\documentclass[conference,compsoc]{IEEEtran}

\usepackage{siunitx}
\usepackage{booktabs}
\usepackage{adjustbox}
\usepackage{subcaption}
\usepackage{graphicx} % for '\includegraphics' macro
\ifCLASSINFOpdf
\usepackage{graphicx}
\else
\fi
\usepackage{algorithm}% http://ctan.org/pkg/algorithms
\usepackage{algpseudocode}% http://ctan.org/pkg/algorithmicx
\algrenewcommand\algorithmicrequire{\textbf{Input:}}
\algrenewcommand\algorithmicensure{\textbf{Output:}}
\usepackage{url}
\usepackage[hidelinks]{hyperref}
\usepackage{caption}
\usepackage{comment}
\usepackage{tcolorbox}
\usepackage{soul,xcolor}
\newcommand{\karthika}[1]{\textcolor{purple}{[(Karthika): #1]}}
\newcommand{\roberto}[1]{\textcolor{red}{[(Roberto): #1]}}

\newcommand{\highlight}[1]{\textcolor{red}{#1}}

\usepackage{xspace}
\newcommand{\seshield}{{\em SEShield}\xspace}
\newcommand{\secrawler}{{\em SECrawler}\xspace}
\newcommand{\senet}{{\em SENet}\xspace}
\newcommand{\seguard}{{\em SEGuard}\xspace}

\begin{document}
\title{SENet: Visual Detection of Online Social Engineering Attack Campaigns}

\author{
%   \IEEEauthorblockN{Michael Shell}
% \IEEEauthorblockA{Georgia Institute of Technology\\
% someemail@somedomain.com}
% \and
% \IEEEauthorblockN{Homer Simpson}
% \IEEEauthorblockA{Twentieth Century Fox\\
% homer@thesimpsons.com}
% \and
% \IEEEauthorblockN{James Kirk\\ and Montgomery Scott}
% \IEEEauthorblockA{Starfleet Academy\\
% someemail@somedomain.com}
}


\author{\IEEEauthorblockN{Irfan Ozen\IEEEauthorrefmark{1},
Karthika Subramani\IEEEauthorrefmark{2},
Phani Vadrevu\IEEEauthorrefmark{3} and
Roberto Perdisci\IEEEauthorrefmark{1}\IEEEauthorrefmark{2} 
}
\IEEEauthorblockA{\IEEEauthorrefmark{1}University of Georgia}
\IEEEauthorblockA{\IEEEauthorrefmark{2}Georgia Institute of Technology}
\IEEEauthorblockA{\IEEEauthorrefmark{3}Louisiana State University}
}


\maketitle

\begin{abstract}
%\boldmath

Social engineering (SE) aims at deceiving users into performing actions that
may compromise their security and privacy. These threats exploit weaknesses in
humans' decision-making processes by using tactics such as pretext, baiting,
impersonation, etc. On the web, SE attacks include attack classes such as
scareware, tech support scams, survey scams, sweepstakes, etc., which can
result in sensitive data leaks, malware infections, and monetary loss. For
instance, US consumers lose billions of dollars annually due to various SE
attacks. Unfortunately, generic social engineering attacks remain understudied,
compared to other important threats, such as software vulnerabilities and
exploitation, network intrusions, malicious software, and phishing. The few
existing technical studies that focus on social engineering are limited in scope
and mostly focus on measurements rather than developing a generic defense. 

To fill this gap, we present \seshield, a framework for in-browser
detection of social engineering attacks. \seshield consists of three
main components: (i) a custom security crawler, called \secrawler, that is
dedicated to scouting the web to collect examples of in-the-wild SE attacks;
(ii) \senet, a deep learning-based image classifier trained on data collected by
\secrawler that aims to detect the often glaring visual traits of SE attack
pages; and (iii) \seguard, a proof-of-concept extension that embeds \senet into
the web browser and enables real-time SE attack detection. We perform an
extensive evaluation of our system and show that SENet is able to  detect new
instances of SE attacks with a detection rate of up to 99.6\% at 1\% false
positive, thus providing an effective first defense against SE attacks on the
web.
    
\end{abstract}
\section{Introduction}
\label{sec:intro}

Social engineering (SE) encompasses a broad spectrum of attacks aimed at
deceiving users into performing actions that may have important negative
consequences for the security and privacy of the users themselves or their
organizations~\cite{mann2008hacking}. These threats exploit weaknesses in
humans' decision-making processes by using tactics such as pretext, baiting,
impersonation, etc.~\cite{KROMBHOLZ2015}. On the web, SE attacks include attack
classes such as scareware~\cite{FakeAVEconomy}, tech support
scams~\cite{miramirkhani2017}, survey scams~\cite{KharrazRK18},
sweepstakes~\cite{clark2013there}, etc., which can result in sensitive data leaks, malware 
infections, and monetary loss. For instance, US consumers loose billions of
dollars annually due to various SE attacks~\cite{FTCReport}.

Unfortunately, social engineering attacks remain understudied, compared to other
important threats, such as software vulnerabilities and exploitation, network
intrusions, malicious software, and botnets. While phishing attacks, which can
be thought of as a specific subclass of SE attacks, have received significant
attention~\cite{PhishingDetectionSurvey}, to the best of our knowledge no
comprehensive defense framework against generic SE attacks has been proposed or
evaluated. 

Previous studies on SE attacks are either limited to discovering and measuring
SE attack campaigns~\cite{Vadrevu_IMC19,WebPushAds} or to studying specific
subclasses of SE
attacks~\cite{miramirkhani2017,FakeAVEconomy,KharrazRK18,Hong2012}.  
Furthermore, existing practical defenses against malicious web pages mostly rely
on reactive approaches that do not focus on the fundamental traits of SE attacks
and tend to lag behind new threats, thus leaving many users exposed to the
latest attack iterations. For instance, URL blocklist services (e.g., Google
Safe Browsing~\cite{GSB}) focus on \emph{where} malicious content is hosted,
rather than detecting and blocking the malicious content itself. Therefore,
attackers evade blocking by simply relocating their attacks to a different
hosting location~\cite{Vadrevu_IMC19}.

A recent work by Yang et al.~\cite{TRIDENT} has started to address the problem
of detecting and blocking web-based SE attacks by proposing an in-browser
defense system named TRIDENT. However, TRIDENT narrowly focuses on detecting
JavaScript code served by low-reputation ad networks that often use SE
techniques, such as transparent overlays, to hijack users' clicks and redirect
them to a potentially malicious page. TRIDENT is able to detect SE
webpages only indirectly,  because low-tier ad network code often (but not
always~\cite{Vadrevu_IMC19}) redirects to SE attacks.  Therefore, there is a
need for more generic solutions that are able to directly detect generic
web-based SE attacks, even when they are not linked specifically to ad network
code.

\vspace{3pt}
\noindent
{\bf Problem Definition and Motivation}:
\label{sec:problemdef}
In this paper, we aim to build an in-browser defense against generic web-based
Social Engineering (SE) attacks, such as {\em fake software downloads}, {\em online sweepstakes}, {\em fake antiviruses}, {\em tech support scams} (TSS), {\em notification permission stealing}, etc.

Although phishing may be considered as a subcategory of SE attacks, we
regard phishing websites as out of scope for this work, due to their different
characteristics, compared to generic SE attacks. Specifically, the reasons for focusing on SE attacks other than phishing are as follows:

\begin{table*}[h]
    \captionof{table}{Examples showing the different visual characteristics of Phishing vs. generic SE attack pages.\label{Tab:sepdif}}
    \centering %this
    \begin{adjustbox}{width=\textwidth} %this
    \small %this    
    \begin{tabular}{@{}cccccc@{}}
    \toprule
    \rotatebox[origin=c]{90}{\textbf{Phishing}} & \raisebox{-.5\height}{\frame{\includegraphics[width=.22\textwidth]{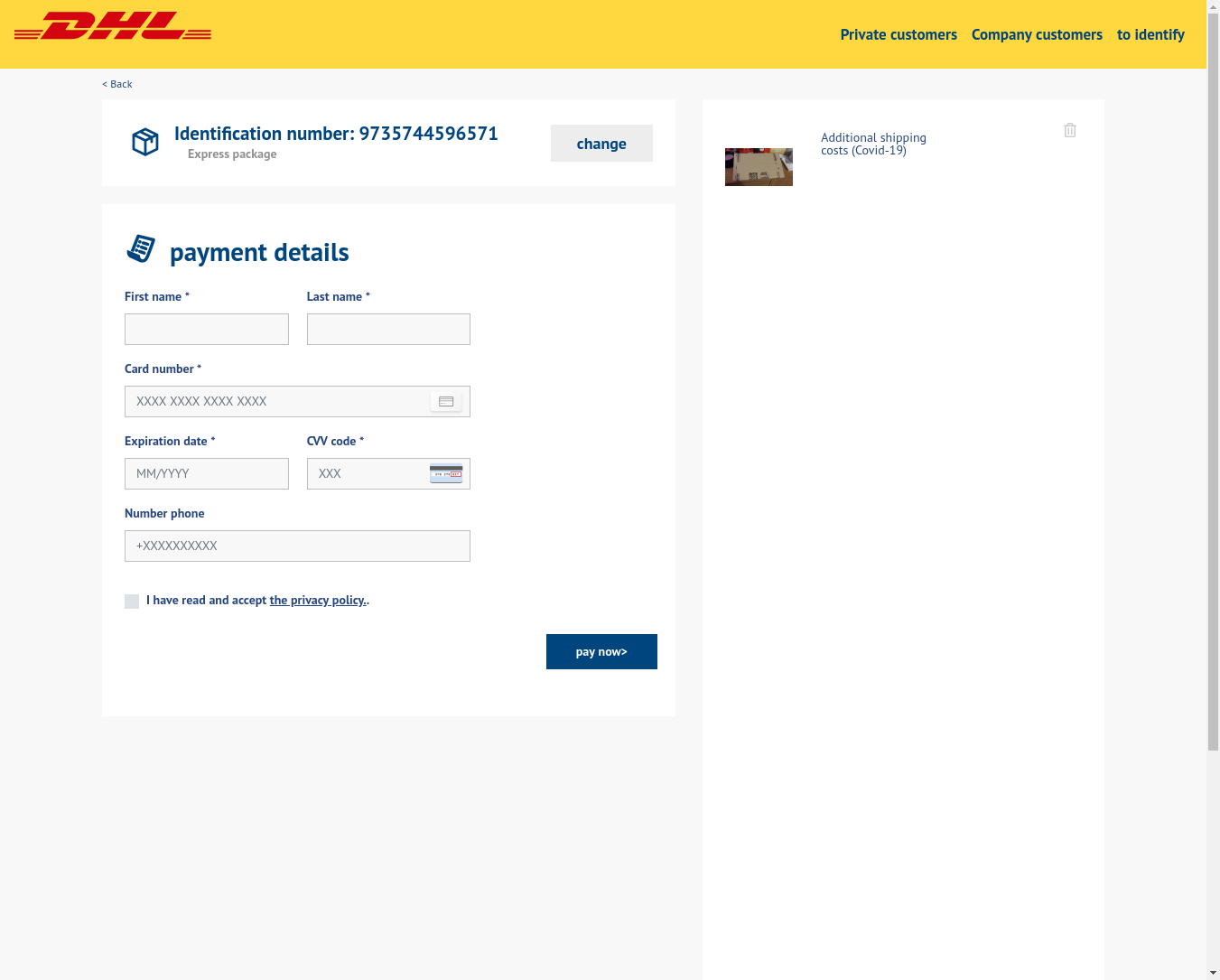}}} & \raisebox{-.5\height}{\frame{\includegraphics[width=.22\textwidth]{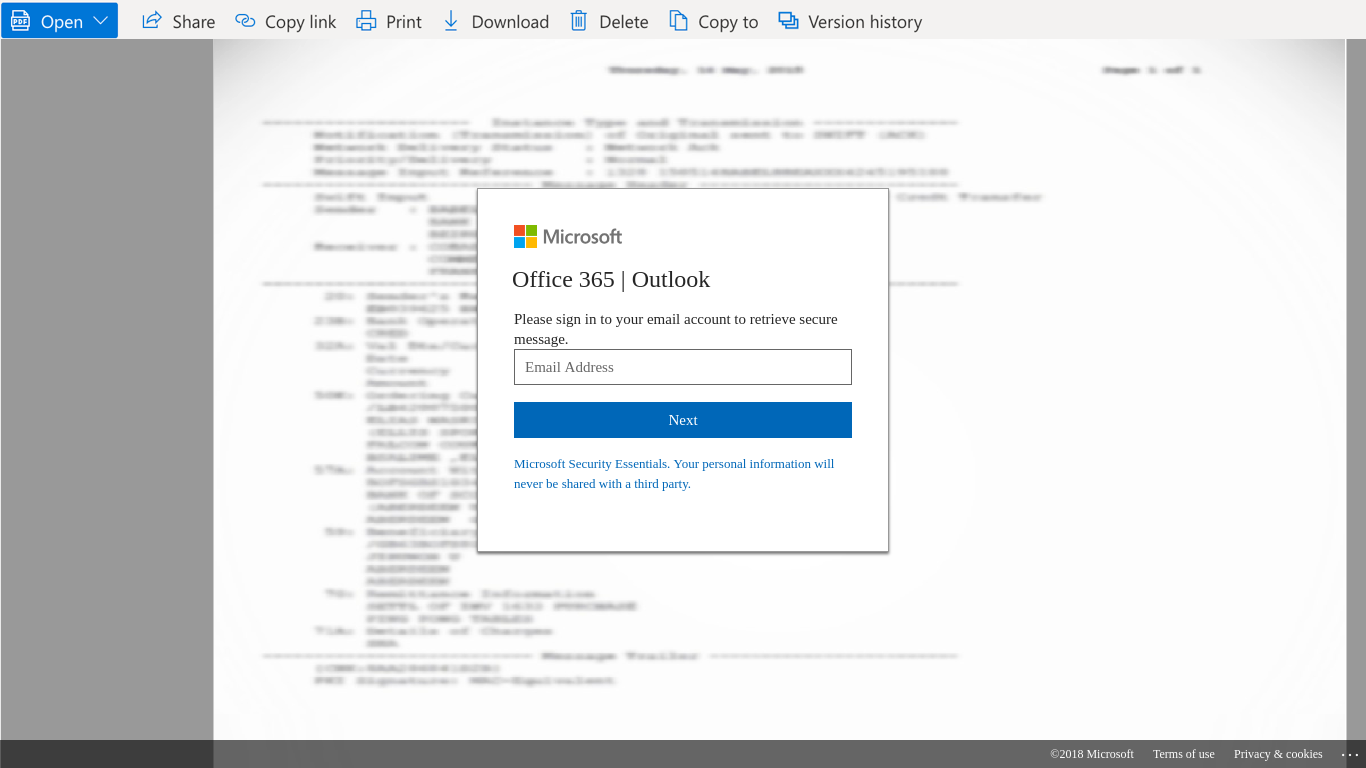}}} & \raisebox{-.5\height}{\frame{\includegraphics[width=.22\textwidth]{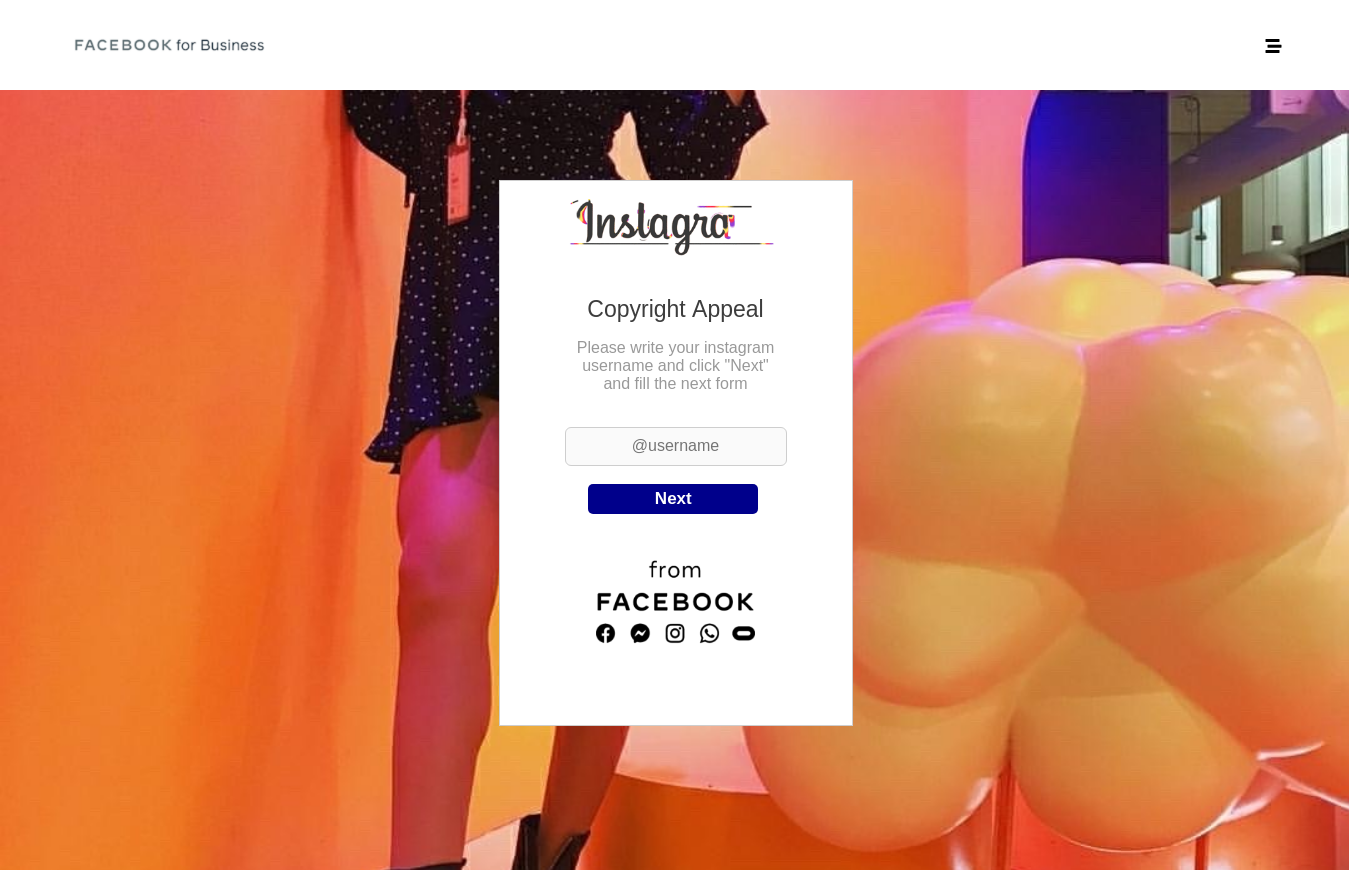}}} & \raisebox{-.5\height}{\frame{\includegraphics[width=.22\textwidth]{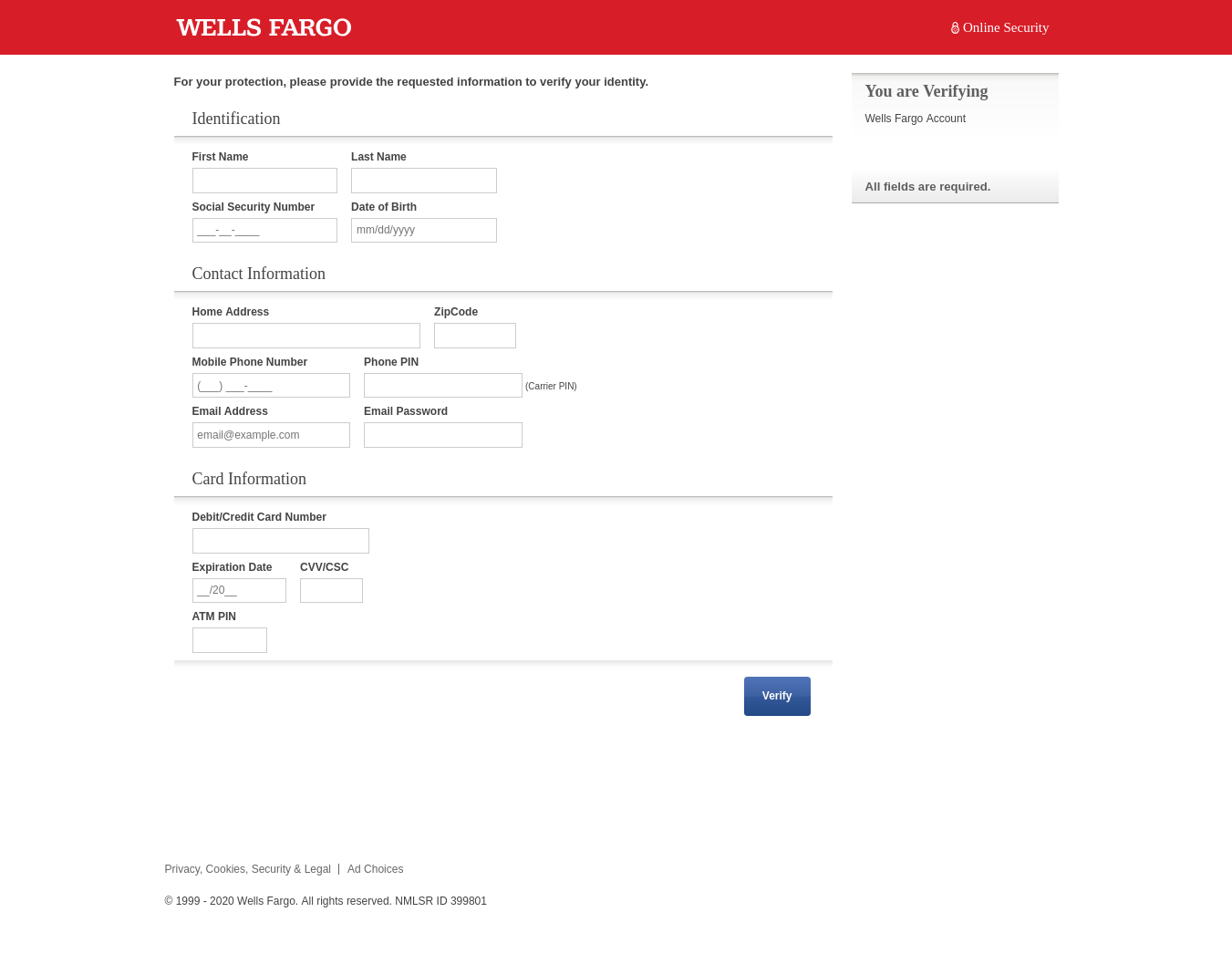}}} \\ \\ \hline \\
    \rotatebox[origin=c]{90}{\textbf{Social Engineering}} & \raisebox{-.5\height}{\frame{\includegraphics[width=.24\textwidth]{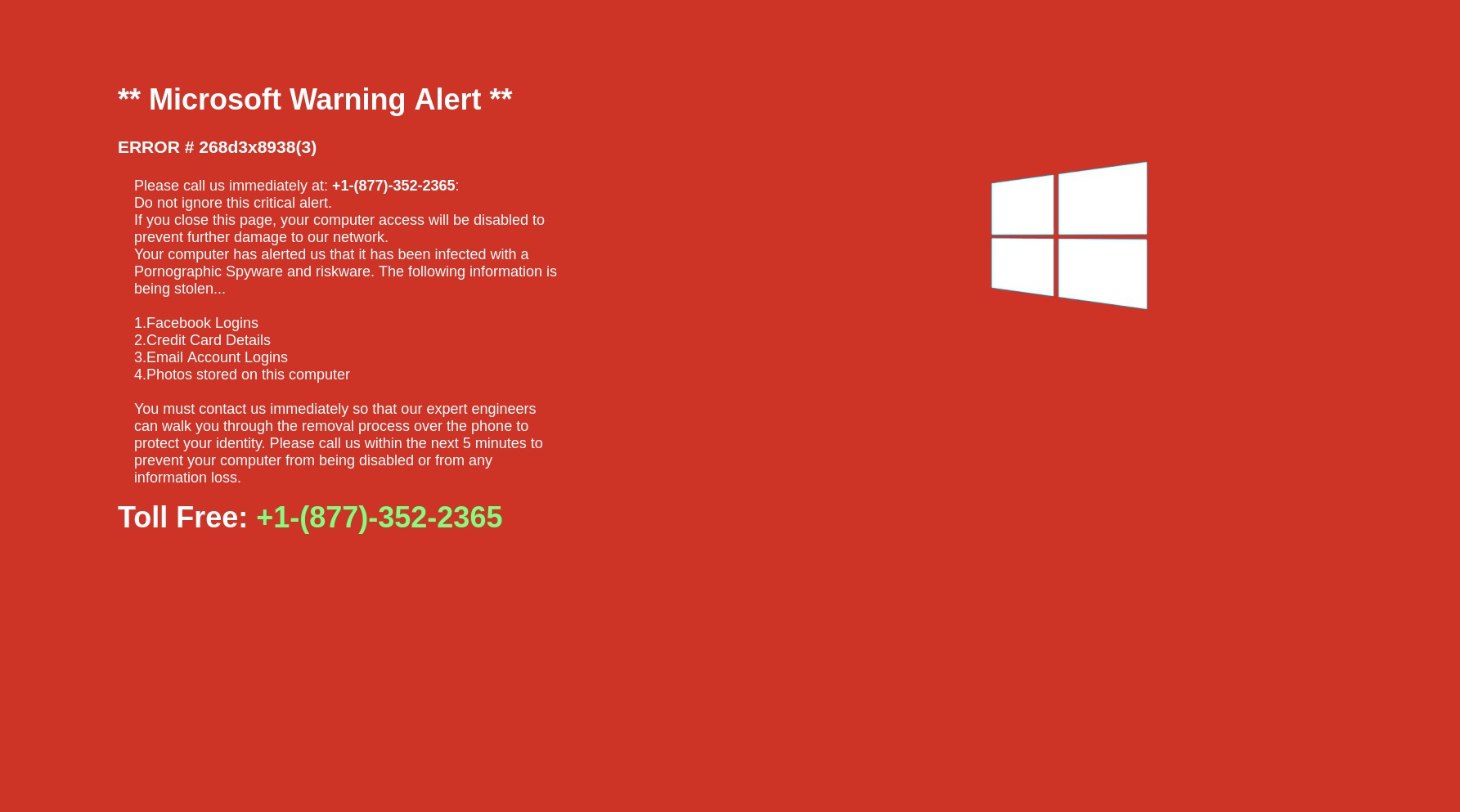}}} & \raisebox{-.5\height}{\frame{\includegraphics[width=.24\textwidth]{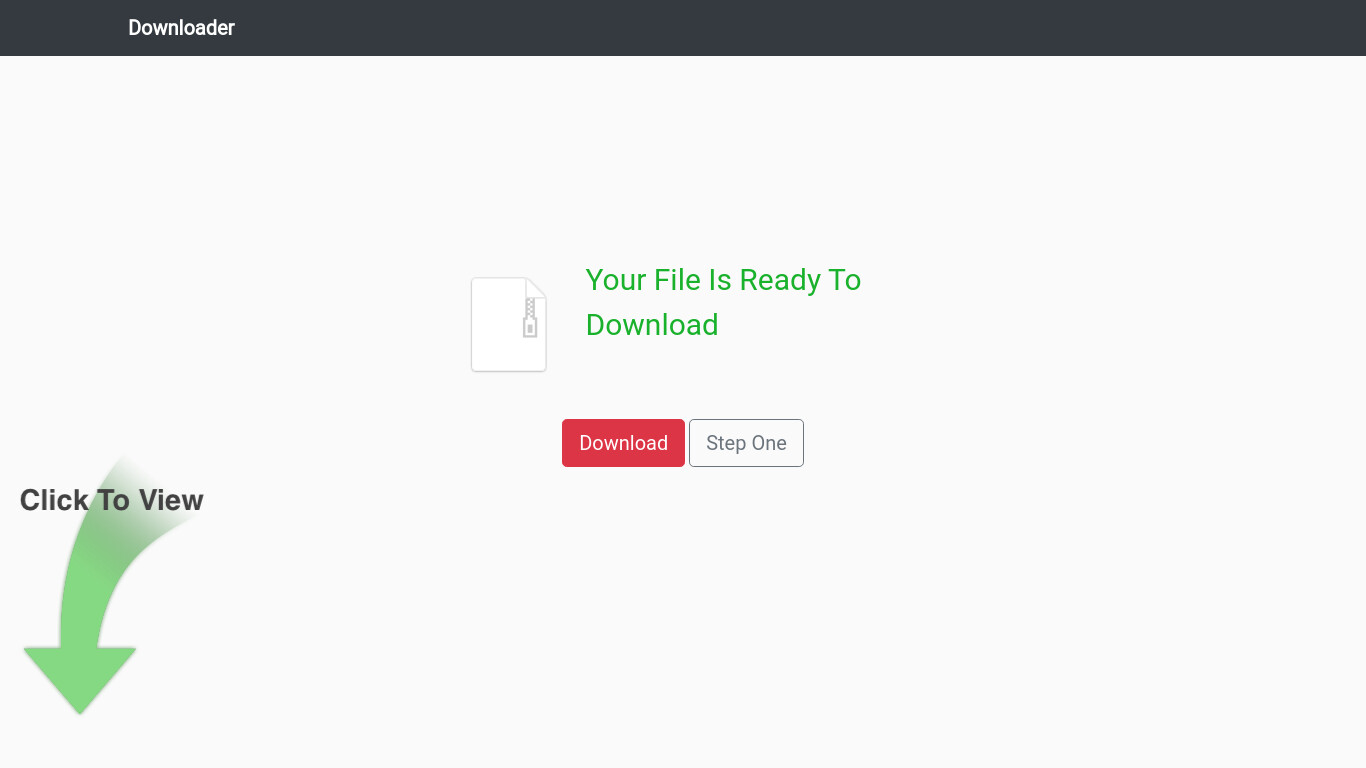}}} & \raisebox{-.5\height}{\frame{\includegraphics[width=.24\textwidth]{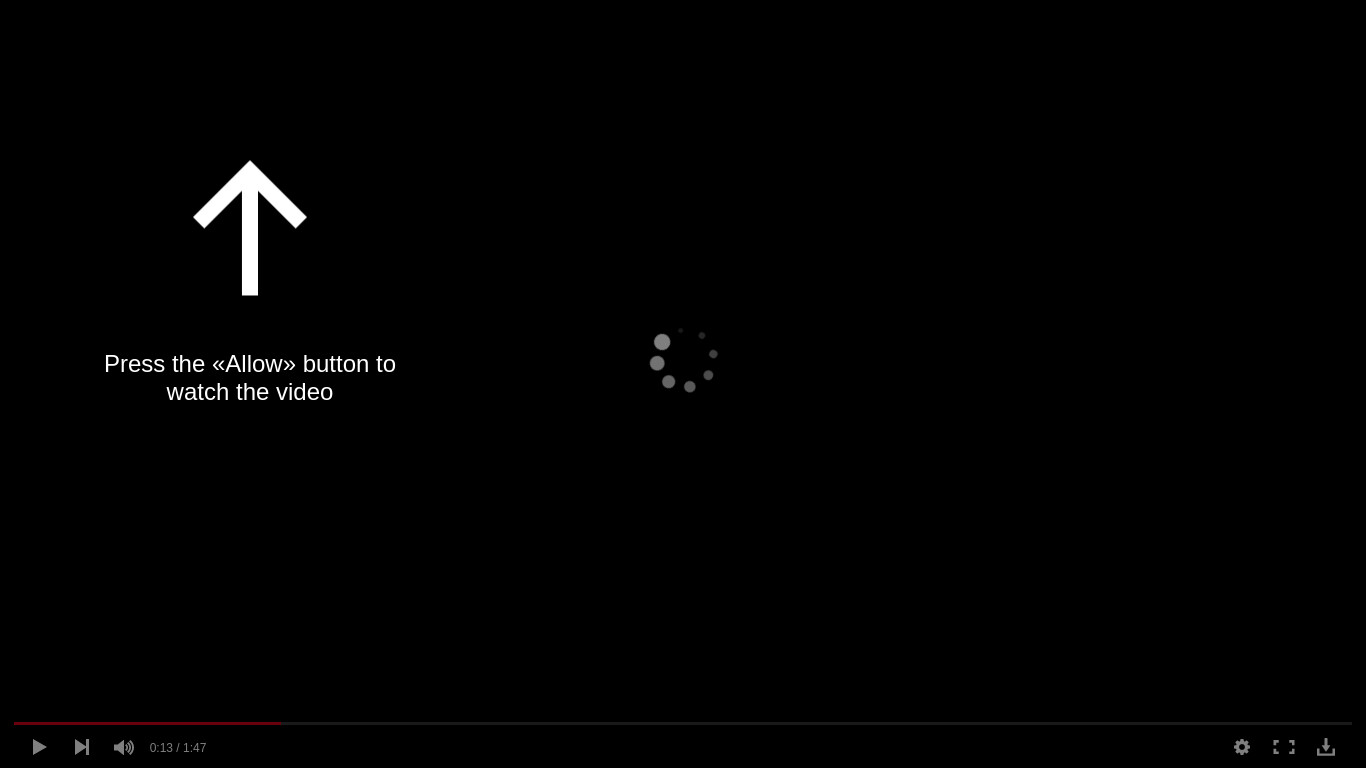}}} & \raisebox{-.5\height}{\frame{\includegraphics[width=.24\textwidth]{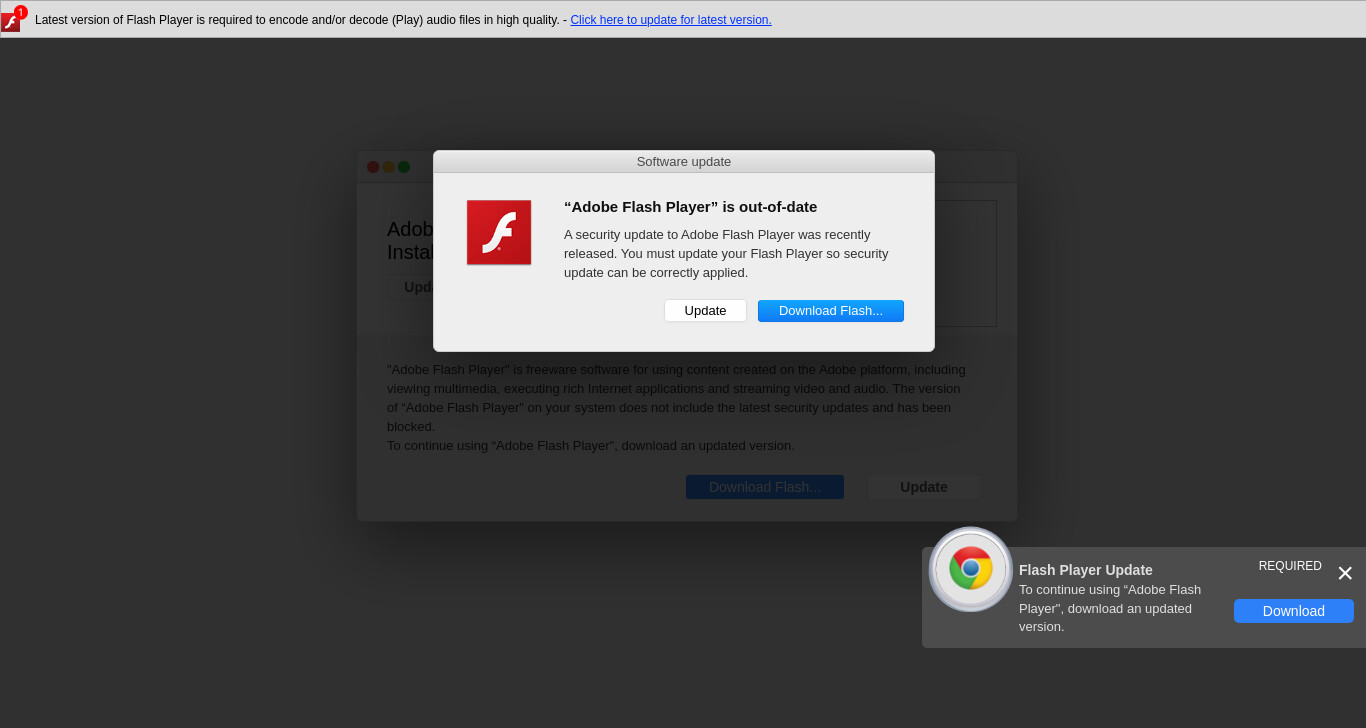}}} \\
    & Tech Support Scam & Malicious Download & Notification Stealing & Fake Software Update \\ \\ \hline 

    \end{tabular}
    \end{adjustbox} %this   
\end{table*}

1) {\em SE attacks differ from phishing}: Phishing attacks typically attempt to
    (i) mimic a specific benign target website (e.g., an online banking site,
    ecommerce site, etc.) and (ii) directly ask users to submit personal
    information, such as login credentials, credit card or social security
    numbers, etc, as part of the attack. On the other hand SE attacks do not
    need to mimic a specific website (though in some cases they may abuse a
    popular brand/logo) and do not need to directly ask users to submit personal
    information (though they may further lead to information stealing). For
    instance, fake software download attacks trick the user into downloading
    malicious software, TSS attacks convince the user to call a fake support
    phone number, notification stealing attacks trick users into granting
    notification permission to potentially malicious websites, etc. The examples
    of real-world screenshots shown in Table~\ref{Tab:sepdif} demonstrate the
    typical visual differences between SE and phishing attacks pages. Notice
    that most SE attacks do not include any input boxes, and that in many
    cases no specific brand or company logo is abused, whereas these properties
    are required in typical phishing pages.

2) {\em Defenses against SE attacks are critically understudied}: While there
    exists an extensive body of work that studies how to detect phishing web
    pages~\cite{PhishingDetectionSurvey}, including using visual
    classification~\cite{VisualPhishNet, lin2021phishpedia, liu2022inferring},
    generic SE attacks remain understudied. Few previous works have addressed
    the problem of defending against web-based SE attacks, with a limited focus
    on collecting/measuring SE campaigns~\cite{Vadrevu_IMC19, WebPushAds} or on
    detecting ad network code that often leads to SE attack
    pages~\cite{TRIDENT}.

\noindent
{\bf Differences Compared to Previous Work}: Visual approaches that have been proposed to detect Phishing web pages~\cite{VisualPhishNet, lin2021phishpedia, liu2022inferring} are specific to Phishing attack characteristics, which are typically not in common with generic SE attacks. For instance, VisualPhishNet~\cite{VisualPhishNet} leverages the similarity between Phishing attack pages and a well-defined list of target benign websites (e.g., banking websites, social media sites, etc.). Phishpedia~\cite{lin2021phishpedia} focuses on detecting the misuse of a target set of logos, whereas PhishIntention~\cite{liu2022inferring} aims to detect the joint appearance of abused logos and credential stealing forms. However, in the case of most SE attacks there are no specific target benign pages that are mimicked, credential stealing forms are typically not present, and while logos are sometimes abused they are often significantly different than benign logos or are completely absent from many SE attack campaigns.

\begin{figure*}[t]
    \centering
    \begin{center}
        % \vspace{18pt}% \caption@setup{format=plain,justification=centering}
        \includegraphics[width=18cm,keepaspectratio]{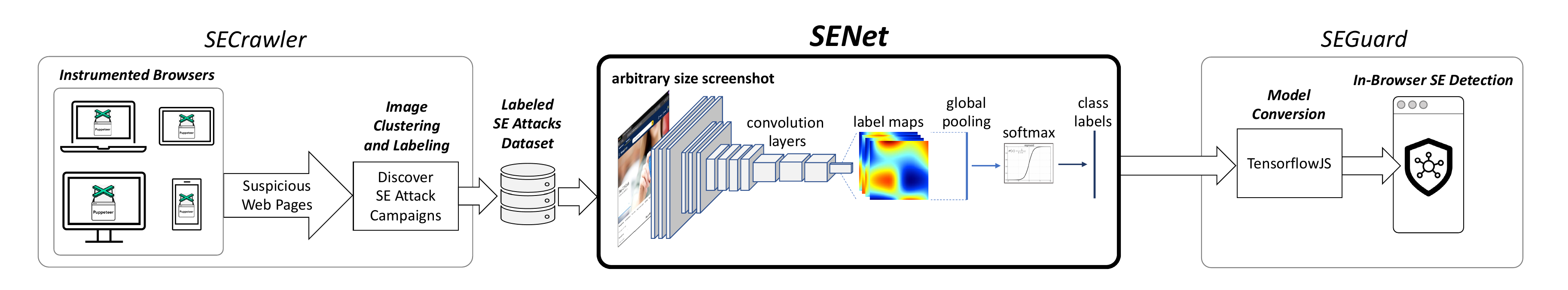}
        \caption{\seshield framework overview. \senet represents the primary contribution and main research focus of this paper. \secrawler and \seguard are implemented as early prototypes that we use for training data collection and to demonstrate the viability of \senet and of the entire \seshield defense framework to enable a practical in-browser defense against SE attacks.}
        \label{fig:overview}
    \end{center}
\end{figure*}

Unlike previous work, we aim to build a SE attack detection framework that
allows us to (i) discover and collect examples of generic SE attacks served by different
SE campaigns, (ii) train a deep learning-based system to visually detect
future variants of SE attacks from observed campaigns, and (iii) deploy the SE
attack detection model into the browser to enable real-time detection of new SE
attack instances (see Figure~\ref{fig:overview}). Additionally, an important
objective of our work is to build a {\em practical} solution that could be
deployed into web browsers running on a variety of devices, including laptop and
desktop computers and mobile devices.

To this end, we focus on detecting web-based SE attacks using a deep learning-based computer
vision approach, which is motivated by the following main observations:
\begin{itemize} 
    \item SE attacks aim to trick users by presenting them with misleading
    content, and thus tend to have a {\em significant visual component} that
    must be present for such attacks to succeed.
    \item Large-scale SE attacks are often organized into {\em attack
    campaigns}, whereby attacks belonging to the same campaign share a common
    theme and thus carry similar content and visual traits.
    \item Recent advancements in deep learning have drastically improved the
    accuracy of image classification tasks, with popular deep learning
    frameworks now also available for languages such as JavaScript, making it
    feasible to build complex in-browser AI systems.
\end{itemize}

% In the next section, we provide an overview of our proposed framework for
% detecting web-based SE attack campaigns.

\begin{comment}
(4) unlike other web-based attacks (such as, browser exploits and cross-site
scripting), SE attacks do not succeed instantaneously, in that they require
humans to process and interact with malicious content, often through multiple
user actions (e.g., multiple clicks to navigate through different pages, typing
on the keyboard, etc.), thus allowing the defender a sufficient compute time
budget for extracting content features, processing the visual rendering of web
pages and web notifications, detecting SE attacks, and alerting the user before
the attacker's ultimate goal is achieved (e.g., personal data is leaked,
malicious software is installed, or a call to a tech support scam is made).
\end{comment}

\noindent
{\bf Contributions}: Motivated by the observations outlined above, in this paper
we make the following main contributions:
% \begin{itemize}
    % \item 

    -- We propose \seshield, a framework to enable the detection of generic
    web-based SE attacks. \seshield consists of three components: (i) a system
    for discovering and collecting examples of SE attack campaigns, named
    \secrawler,  (ii) a deep-learning based visual classifier called \senet, and
    (iii) a browser extension named \seguard that classifies web pages in real
    time and alerts the user when an SE attack is detected. Among these three
    components, most of our research efforts and contributions are dedicated to
    designing, implementing and evaluating the \senet classifier. At the same
    time, we also build a prototype of \secrawler and \seguard to enable the
    collection of SE attacks needed to train \senet and to demonstrate the
    viability of the \seshield defense framework.
    
    % \item 
    -- We implement \secrawler by extending previous work~\cite{Vadrevu_IMC19}
    to discover and collect fresh examples of SE attack campaigns. We then label
    the collected data by employing three different human labelers
    and using a systematic labeling strategy to obtain high-quality ground truth, which we use to train and evaluate the \senet
    classifier. Our dataset consists of 7,484 labeled SE attack pages (captured using 30 different emulated devices and screen sizes) that belong to 74 different SE attack campaigns. We will release our dataset to the research community to enable further research in this area.
     
    % \item 
    -- We design \senet as a visual detection model that enables the
    classification of images of arbitrary sizes, and develop a new distributed training approach to be able to train our model with limited GPU resources. \senet can be deployed in the
    browser and is able to visually detect SE attack webpages independently of
    the size of the screen (or browser window) where the page is rendered.
    
    % \item 
    -- We conduct an extensive evaluation of \senet to demonstrate its ability
    to detect new instances of SE attacks linked to campaigns observed by
    \secrawler. We show that \senet can accurately detect new SE attack instances 
    with up to 99.6\% recall at 1\% false positive rate, that it
    is able to generalize to input images (i.e., webpage screenshots) related to
    previously unseen screen sizes, and that it is also able to generalize to
    detecting a number of never-before-seen SE campaigns.
    
    % \item 
    -- We implement a prototype of \seguard as a browser extension using
    TensorFlowJS~\cite{TensorFlowJS} to demonstrate the utility of the entire
    \seshield framework as a practical defense against SE attack webpages. We plan to release all components of our \seshield framework as open-source
    software.
   
    \begin{comment}
    % \item 
    -- Additionally, we will release the dataset of labeled SE attack
    campaigns collected with \secrawler, which consists of \highlight{30}
    campaigns and more than 3000 SE attack screenshots and 263 metaclusters taken on 30
    different emulated devices and screen sizes. To the best of our knowledge,
    this will be the largest dataset of real-world SE attack webpages available
    to the research community. \roberto{These numbers need to match the experimental setup numbers, no? They currently don't.}
    \end{comment}
% \end{itemize}

\section{Framework Overview}
\label{sec:sysoverview}
In this section, we present an overview of our SE defense framework, named
\seshield, which aims to offer a practical defense against generic web-based SE
attacks. 
Figure~\ref{fig:overview} shows the three components that make up our framework:
\secrawler, \senet, and \seguard (we introduce each component in more detail
below). In this paper, we focus most of our research efforts on \senet, and
build an early prototype of \secrawler and \seguard to enable the collection of SE
attack examples and to demonstrate the viability of \senet and of the entire
\seshield defense framework.

As mentioned in Section~\ref{sec:problemdef}, we focus on detecting SE attacks
beyond phishing, such as {\em fake software downloads}, {\em scareware}, {\em
sweepstakes}, {\em tech support scams}, etc. To detect SE attacks, we take a
deep learning-based computer vision approach to leverage the often glaring
visual cues exhibited by SE attack campaigns. Additionally, to enable a
practical deployment on devices with a variety of screens and browser window
sizes, we design \senet to be able to classify images (webpage screenshots) of
arbitrary sizes.

\subsection{Approach Motivation}
At a high level, our approach towards building an SE detection system is informed by the following observations:

{\em SE attacks have a very significant visual component}: SE attacks exploit a
victim' decision-making processes by using a number of deception and persuasion
tactics. To attract users' attention and deceive/persuade them of the veracity
of an invented scenario (e.g., the user's device is compromised, a software
update is needed to watch a video, etc.), web-based SE
attacks make use of strong visual cues, including fake dialog boxes, abused
logos/brands, flashy messages, etc., as shown by the examples in Table~\ref{Tab:sepdif} and Figure~\ref{fig:MCLUSTERS}. Therefore, our image classification approach aims to detect the
visual traits of such attacks.

{\em SE attacks are typically orchestrated via SE campaigns}: To scale the
attacks and facilitate their distribution to a large number of potential
victims, SE attacks are orchestrated via SE campaign. Attack campaigns share
some similarities with legitimate advertisement campaigns. In legitimate
advertisement, the same product/image (or few image variants) is advertised in
multiple venues (either physical or online). Similarly, an SE campaign aims to
``advertise'' visually similar/same SE attacks on multiple different
websites, as also noted in~\cite{Vadrevu_IMC19}. Thus, it may be sufficient to
observe and learn from few examples of an SE campaign to allow us to detect
future instances of the same or similar SE attacks on different websites.

Based on the above observations, we build a system that aims to continuously collect examples of in-the-wild SE attacks belonging to a variety of campaigns and train a model that can detect new SE attack variants that share visual traits with previously observed SE campaigns.

\subsection{Framework Components}
At a high level, the three components of our \seshield framework work as
follows:

\vspace{1pt}
\noindent
\secrawler: As we aim to learn to automatically identify web-based SE attacks,
    we first need to be able to continuously collect fresh examples of web pages
    related to SE attack campaigns. To this end, we reimplement and extend the
    SE attack discovery approach proposed by Vadrevu et
    al.~\cite{Vadrevu_IMC19}. Specifically, we extend~\cite{Vadrevu_IMC19} by
    significantly increasing the number of different supported devices and
    screen resolutions (from 5 to 30) and by keeping track of the path followed
    by the crawler to reach a potential SE attack page. We then use these
    improvements over~\cite{Vadrevu_IMC19} to collect larger and more detailed
    datasets of SE attack campaigns, which we use for training and evaluating
    our \senet module. Our data collection and systematic labeling process will
    be presented in Section~\ref{sec:secrawler}. 

\vspace{3pt}
\noindent
\senet: The \senet module represents the primary contribution and main focus of
    our research. Using examples of SE attack campaigns collected with our
    \secrawler, as well as negative examples consisting of benign web pages, we
    train a novel deep learning-based visual SE attack classifier. As mentioned
    earlier, we aim to deploy our SE attack classifier on a variety of devices,
    including laptops, desktops, and mobile devices, which can have widely
    different screen resolutions. Therefore, we want our classifier to support
    input images (i.e., web page screenshots) of arbitrary sizes. However, this is not always easy to
    accomplished with traditional convolutional neural networks (e.g., VGG,
    ResNet, etc.) and training approaches, which are designed to expect images
    of fixed size in input. To solve this issue, we adapt existing deep learning
    architectures to build an image classifier that can generalize to
    classifying web page screenshots of any screen size with high accuracy. In
    addition, we design a new distributed learning approach inspired by
    federated learning to be able to train our model over images of arbitrary
    sizes with limited GPU resources. We will present the details of the
    architecture and implementation of \senet in Section~\ref{sec:senet}.

\vspace{3pt}
\noindent
\seguard: Once our \senet classifier is trained, we demonstrate the practicality
    of our SE defense framework by developing a prototype Chrome extension that
    deploys the {\em SENet} model into the browser. Our browser extension, named
    {\em SEGuard}, is programmed to detect SE attacks in near real time. It does
    so by capturing a screenshot of each webpage on which a meaningful user
    interaction is observed, such as a mouse click or key press, and by using
    {\em SENet} to classify the screenshot image as either {\em SE attack} or
    {\em benign} (i.e., not SE-related). Note that our system could also be integrated natively into
    the browser's code, similar to Chrome's built-in phishing-specific
    detector~\cite{google_phishing}. More details on the implementation of {\em
    SEGuard} are discussed in Section~\ref{sec:seguard}.

In the following sections, we describe the components of our system in more detail and present a comprehensive evaluation of the entire \seshield framework, with particular focus on measuring the detection performance and generalization abilities of our \senet classifier.

\section{Framework Details}
\label{sec:details}
In this section, we provide more details on each of the components of the \seshield framework, with particular focus on \senet, which represents our main contribution. 

\subsection{SECrawler}
\label{sec:secrawler}

While research on phishing defenses can leverage URL feeds such as PhishTank~\cite{phishtank} and OpenPhish~\cite{openphish} to collect ground truth data, to our best knowledge there exist no analogous feeds or repositories that would allow us to readily collect fresh samples of generic SE attacks. Therefore, we had to build our own system for discovering and collecting recent examples of such attacks in the wild. To this end, we extended the SE attack campaigns discovery approach proposed in~\cite{Vadrevu_IMC19}. Briefly, the approach is based on a {\em crawler farm} that is seeded with URLs that are likely to lead to SE attacks. A crawler consists of an instrumented browser that loads a seed URL and automatically interacts with web pages in an attempt to trigger a redirection to an SE attack. A number of heuristics are used to pilot the browser (please refer to~\cite{Vadrevu_IMC19} for details) to collect {\em landing pages} that have a high likelihood of being related to SE attacks. As these landing pages are visited, the crawler records the related URL and a screenshot. Then, to discover SE attack campaigns and filter out noise (i.e., non-SE pages), a webpage screenshot clustering algorithm is used, which is based on the intuition that SE attacks are typically orchestrated into SE campaigns that distribute the same (or very similar) attacks across multiple malicious domain names.

To enable our data collection, we reimplemented an instrumented browser similar to the one proposed in~\cite{Vadrevu_IMC19} using Puppeteer~\cite{puppeteer} and extended it to add the following improvements: 

\begin{comment}
perceptual clustering of screenshots and then filter these clusters to isolate website groups that host \textit{visually similar content} on \textit{different domain names}.  This relies on the intuition that SE attacks are commonly organized as part of multi-site campaigns in order to evade domain-based take-down efforts from the industry.  Another key idea is to run the crawler specifically on seed websites discovered to be using low-tier,  unscrupulous ad networks which commonly act as fertile grounds for distributing SE attack content.  The authors show that their proposed approach is effective as it leads to clusters that are mostly (about 83\%) related to generic SE attack content. Therefore, we relied on this approach for collecting ground truth data for training our \senet module. 

We used Google Chrome's Puppeteer framework~\cite{puppeteer} to build \secrawler. We also made use of this opportunity to make some improvements over prior work to improve the quality ground truth SE attack data that we can collect.  Below,  we will briefly describe these improvements.  Later on,  we will give more details about the data that \secrawler collected. 
\end{comment}

% \vspace{3pt}
\noindent \textit{Improved crawler heuristics:} In~\cite{Vadrevu_IMC19}, ``click-worthy'' page elements were prioritized by sorting them simply in decreasing element size order (i.e., based solely on the size of an element on the rendered webpage). To allow \secrawler to more quickly reach SE attack content, we empirically derived additional heuristics. For instance, we prioritize interacting with page elements (e.g., buttons, images, links, etc.) that include keywords such as {\em update}, {\em download}, {\em play}, etc. (we used a total of 38 keywords often found on pages that lead to SE attacks). This small improvement over~\cite{Vadrevu_IMC19} allowed our crawler to reach more attacks. Additionally, compared to~\cite{Vadrevu_IMC19}, we added the use of~\cite{stealth} to improve the ``stealthiness'' of our crawlers (i.e., make it more difficult for pages to detect browser automation) and yield a larger variety of SE attacks.
\begin{comment}
Seeing the many challenges that~\cite{Vadrevu_IMC19} faced when trying to appear stealthy and evade triggering ad networks' evasion techniques,  we leveraged a systematically developed stealth library for Puppeteer \cite{stealth} in our crawler.  
\end{comment}

% \vspace{-3pt}
\noindent \textit{Improved data granularity:}
More importantly, compared to~\cite{Vadrevu_IMC19}, we collect finer-grained contextual information around SE attacks with two improvements: (1) We implemented our \secrawler to track all steps that a crawler followed to reach a given page, which enables {\em attack traceback}; 
% In later sections, we will discuss how this additional traceback capability helped us develop more effective SE attack detection models. 
(2) Because one of our main goals is to develop a practical defense that can be deployed on devices with widely different screen resolutions, we significantly extended the number of devices and screen sizes emulated by our crawlers, from 5 to 30; To select the screen sizes and devices to be emulated, we relied on external statistics~\cite{statcounter-resolutions} as well as results from a prior IRB-approved user study (from an unrelated research project) involving 400 \url{MTurk.com} users that allowed us to derive the most popular viewport sizes and screen aspect ratios to be embedded in our \secrawler system.

% \vspace{-3pt}
\noindent \textit{\phantomsection\label{labeling1}Systematic labeling process:} As in~\cite{Vadrevu_IMC19}, after suspected SE attacks are clustered into potential campaigns and noise is removed using conservative heuristics, there is a need for manual intervention to confirm and label SE attack campaigns. Unlike in~\cite{Vadrevu_IMC19}, we use a systematic labeling approach. Namely, we label potential SE attack campaigns and remove non-SE pages by using three different human labelers with expertise on web security and social engineering attacks who were tasked with labeling each cluster of webpage screenshots as either {\em SE Campaign} or {\em benign} (i.e., not SE).  
%
\begin{comment}
\begin{enumerate}
    \item \textbf{Strictly SE(SSE)}: Web pages that state an obvious lie to trick users to take unwanted or urgent actions such as installing software, giving permissions, entering their credentials
    \item \textbf{Likely SE(LSE)}: Web pages with no obvious lie but still contains different indicators such as 1) design similar to previously studied SE attacks or 2) a cluster with images of similar designs with slight modifications in their logo or brand name
    \item \textbf{Other}: Any other web pages that do not fit the above criteria. 
\end{enumerate}
\end{comment}
%
After the labelers independently completed their tasks, they met to discuss and
attempt to resolve possible label disagreements. In the rare cases in which
consensus on the label to be assigned could not be found, they marked the
related cluster as {\em unknown} and excluded it from further consideration.
This systematic data labeling process allowed us to collect higher quality
SE attack examples. We computed the inter-rate reliability score using Krippendorff's alpha score~\cite{krippendorff2011computing} among the labelers and obtained $\alpha=0.82$, which indicates a high level of agreement~\cite{de2012calculating}.
Besides collecting and labeling examples of SE attacks, we also used our \secrawler to collect benign webpages under 30 different screen resolutions by seeding it with a list of popular websites (the top 600 websites according to Tranco~\cite{tranco}).

\subsection{SENet}
\label{sec:senet}

To provide an effective solution for in-browser detection of generic SE attacks, we need to take into account that web pages can be rendered on devices with widely different screen resolutions or browser windows of arbitrary size. Therefore, unlike traditional image classification models, which are typically trained assuming input images of fixed size, our SE detection system needs to be able to classify images of arbitrary size and generalize to different types of SE attacks.

To build \senet in a way that enables it to classify images of arbitrary sizes,
we proceed as follows. We start from popular pre-trained deep learning models
(e.g., VGG19, ResNet50 Xception, etc.), to take advantage of fine-tuning
approaches and avoid having to train a large model from scratch. We adapt the model's architecture by removing their
pre-trained classifier ``head'' and replacing it with a new binary classification layer. Furthermore, we replace the input layer to remove the hard constraint on the expected size of input images (we set the input hight and width to ``None''). We then
fine-tune the model by training it on our dataset of SE attacks and benign
pages. Finally, we compare the performance of different models and choose the
model that provides the best trade-off between true and false positives (details are provided in Sections~\ref{sec:senet_setup} and~\ref{sec:expresults}). 

Some models, such as VGG16, VGG19, etc., include a flattening layer in their classifier ``head,'' to reshape the output of convolutional layers into a vector that can be used in input to dense layers. However, the flattening layer is expected to produce a fixed size output vector, which is not the case if the input image is not of the specific fixed size expected by the model. Thus, we adapt such models by removing the classifier ``head'' and flattening layer, and by adding a global pooling layer, whose output size is independent of the input image size. As a concrete example, consider the VGG19 model pretrained on ImageNet~\cite{Imagenet}, which can only
accept images of a fixed size 224x224, as shown in Figure~\ref{fig:vgg19model}. The main obstacle to adapting
VGG19 to classifying images of arbitrating size is the Flatten layer, which is
generates a vector of fixed size  (25,088 elements), and the subsequent Dense
layers. To adapt the model, we remove
the last 4 layers, add a global pooling layer, and then include a binary dense
layer to predict {\em SE} vs. {\em benign} web page screenshots. Finally, we fine tune this
new model with training examples of SE attacks and benign web pages.

\begin{figure}
\small{a) Original VGG19 architecture}
\begin{tiny}
\begin{verbatim}
 Layer (type)                Output Shape              Param #   
=================================================================
 input_1 (InputLayer)        [(None, 224, 224, 3)]     0         
 block1_conv1 (Conv2D)       (None, 224, 224, 64)      1792      
 ....   
 block5_conv4 (Conv2D)       (None, 14, 14, 512)       2359808   
 block5_pool (MaxPooling2D)  (None, 7, 7, 512)         0         
 flatten (Flatten)           (None, 25088)             0        
 fc1 (Dense)                 (None, 4096)              102764544  
 fc2 (Dense)                 (None, 4096)              16781312
 predictions (Dense)         (None, 1000)              4097000   

\end{verbatim}
\end{tiny}

\small{b) Adapted VGG19 architecture for arbitrary size input images}
\begin{tiny}
\begin{verbatim}
 Layer (type)                Output Shape              Param #   
=================================================================
 input_1 (InputLayer)        [(None, None, None, 3)]   0         
 block1_conv1 (Conv2D)       (None, None, None, 64)    1792      
 .... 
 block5_conv4 (Conv2D)       (None, None, None, 512)   2359808   
 block5_pool (MaxPooling2D)  (None, None, None, 512)   0         
 gmp2d_1 (GlobalMaxPooling2D)(None, 512)               0         
 predictions (Dense)         (None, 2)                 1026      
\end{verbatim}
\end{tiny}
\caption{Original vs. adapted VGG19 architecture.}
\label{fig:vgg19model}
\end{figure}

Notice that, in our case, simply resizing the input images to fit a fixed input
size is not a good option. The reason is that, as mentioned earlier, web page
screenshots can have widely different sizes, depending on the device and browser
in which they render. Thus, resizing
an input screenshot can sometimes cause significant distortions,
% of the image's content, as shown in \highlight{Table~\ref{Tab:sizedif}},
which can make accurate classification difficult. Instead, our approach
allows us to keep the aspect ratio of all input web pages intact and at the same
time to achieve high classification accuracy. 
% \roberto{Add example showing that resizing web pages to 224x224 can cause significant distortion}.

%%%% Removed because this particular examples do not show a lot of distortion like others %%%%
\begin{comment}
\begin{table}[h]
    \captionof{table}{Examples of resized and original images.\label{Tab:sizedif}}
    \centering %this
    \begin{adjustbox}{width=\columnwidth} %thisis
    \small %this    
    \begin{tabular}{@{}ccc@{}}
    \toprule
    \rotatebox[origin=c]{90}{\textbf{224x224}} & \raisebox{-.5\height}{\frame{\includegraphics[width=.15\textwidth]{figures/2resize.png}}} & \raisebox{-.5\height}{\frame{\includegraphics[width=.15\textwidth]{figures/6resize.png}}} \\ \\ \hline \\
    \rotatebox[origin=c]{90}{\textbf{Original Size}} & \raisebox{-.5\height}{\frame{\includegraphics[width=.10\textwidth]{figures/1resize.png}}} & \raisebox{-.5\height}{\frame{\includegraphics[width=.24\textwidth]{figures/5resize.png}}}\\ \hline 

    \end{tabular}
    \end{adjustbox} %this   
\end{table}
\end{comment}

Also, one may think that we could address the challenges mentioned earlier by building a separate classifier for each of the possible input image sizes. However, this approach would incur three major problems: (i) it is difficult to collect a large enough number of training examples for each different screen resolution, (ii) on laptop and desktop devices, the browser window could be set by the user to any arbitrary size and aspect ratio that cannot be easily anticipated, and (iii) training separate classifiers would make it difficult to leverage similarities between web pages (malicious or benign) rendered on different screen sizes. Therefore, we chose to train a single model that leverages training images of many different sizes, which helps us to better generalize to any arbitrary input size at test time.

\vspace{3pt}
\noindent
{\bf Model selection process}: To find a pre-trained model that performs well on our
specific web page classification problem, we adapted (when needed) and compared
five different popular models that are characterized by different model
complexity (i.e., number of weights) and accuracy. Namely, we fine-tuned and
compared VGG-16 and VGG-19~\cite{simonyan2015deep},
Inception-ResNet-v2~\cite{Szegedy}, ResNet50V2~\cite{HeZRS16}, and
Xception~\cite{Chollet}, all pre-trained on the
ImageNet~\cite{BiancoCCN18} dataset.

To reduce the amount of compute resources and time needed for the comparison of
these five models with multiple combinations of hyperparameters, we trained each
of the models on a fixed reduced-size dataset of 500 real-world SE
attacks and 500 benign images and tested on 100 benign and 100 SE images. During these preliminary model
selection experiments, we found that VGG-19 performed about 20\%  better than
other models in terms of both false positives and false negatives. We therefore
selected VGG-19 as the reference model for all subsequent experiments, whose
results are reported in Section~\ref{sec:expresults}.

\vspace{3pt}
\noindent
{\bf Implementation challenges and solutions}: One of the main implementation
challenges we faced while working with models that support images of arbitrary
sizes is related to distributing the training workload. First, our dataset of
web page screenshots is too large to fit in a single GPU's memory (we
experimented with 48GB A40 NVIDIA GPUs). Furthermore, while popular deep learning
frameworks support distributed training (e.g.,
Tensorflow~\cite{DistributedTensorflow}), the distributed training APIs typically expect
input tensors of consistent, fixed size. After several unsuccessful attempts to
adapt existing frameworks, we then realized that this would be very challenging
for a small research team and we then chose to instead build a custom
distributed training framework inspired by federated learning~\cite{McMahan}
that would fit our needs.

Federated learning \cite{McMahan} allows for training a model based on data that
is distributed across numerous devices, whereby each device performs local
training and only shares model updates with a central server. The central server
then aggregates the model updates to obtain a final model. 
To train \senet, we devise an approach inspired by federated learning that we
named {\em localized federated learning}, in which we create a set of virtual
devices, or training {\em clients}, and a training {\em server}, and assign each
client with a different portion of the training dataset. Then, each client
trains a model on the data assigned to it and sends the model weights to a the
training server, which computes a global trained model. This strategy allows us
to avoid loading the entire dataset at once and exhaust the GPU's memory. Also, it
allows us to partition the dataset in small sub-datasets and to assign
randomly selected sub-datasets to different clients.  

Since each
client can train its local model separately from the others, we only need to train
the model on a sub-dataset at a time, thus avoiding GPU memory exhaustion
problems. Additionally, this makes it easy to scale to multiple GPUs, as
different clients can train in parallel on different GPUs. It is also worth noting that each client can train on images of different sizes, which helps with generalization. The only constraint is that training images in a single minibatch must all be of the same size. However, during an epoch each client will be trained on many minibatches belonging to its assigned sub-dataset, with each minibatch being related to a randomly selected set of images all having the same size within that minibatch.
Algorithm~\ref{alg:ALG1} summarizes the high-level steps involved in training
\senet using this approach.

\begin{algorithm}
    \small
    \caption{Training \senet with localized federated learning}
    \label{alg:ALG1}
    \begin{algorithmic}
    \Require {\\
        $M:$ VGG19 pre-trained on Imagenet\\
        $GE:$ Global epochs\\
        $CE:$ Client epochs \\
        $CC:$ Client count\\
        $SW:$ List of client weights}
    \Ensure {A fine-tuned VGG19 model}
    \\
    \end{algorithmic}
    \begin{algorithmic}[1]
    \ForAll  {$j=1$ to $GE$}
        \ForAll  {$i=1$ to $CC$}
            \State        Set {\em clientModel}.weights = $M$.weights 
            \State        Assign dataset partition to the $i$th client
            \State        Group images in $i$th dataset partition into minibatches based on image size
            \State        Train the {\em clientModel} for $CE$ epochs
            \State        Scale the weights of {\em clientModel} based on $CC$
            \State        $SW$.insert({\em clientModel}.weights)
        \EndFor 
        \State Set {\em averageWeights} = average of weights in $SW$
        \State Set $M$.weights = {\em averageWeights} 
        \State Evaluate $M$ on validation data
    \EndFor 

    \end{algorithmic}
\end{algorithm}
\subsection{SEGuard}
\label{sec:seguard}

To further demonstrate the practicality of our system, we develop a
proof-of-concept extension for the Chrome browser that allows for detecting SE
attacks in real time. As the user browses the web, every time a new URL is
visited, the extension waits to check if the user interacts with the page (e.g.,
via mouse click). Then, if an interaction occurred, it takes a screenshot and
passes the related image to the \senet classifier to determine if the visited
page contains an SE attack, and if so it presents the user with an alert. 

More precisely, \seguard takes the following steps. (i) When a new page is visited, we can first check if the associated domain name is present in an allowlist (e.g., a list of the top most popular websites according to Tranco~\cite{tranco}), in which case no action is taken. (ii) If the page is not in the allowlist, \seguard waits for a meaningful user interaction with the page, such as a {\em click}, {\em keypress}, etc., event. If one of such events occurs, a screenshot of the browser viewport is captured. Waiting for a meaningful user interaction with the page is motivated by the fact that, unlike drive-by browser exploits, SE attacks often require the user to interact with the page to click on a button, click on a (fake) notification, etc., for the attack to succeed (notice that other heuristics may also be used to determine if/when to take a screenshot). (iii) Once a screenshot is taken, we feed into our \senet model, and if it is classified as an SE attack the user is informed via an alert drawn over the page by the extension.

As mentioned earlier, to classify a page we leverage \senet's
deep learning model. However, we first need to adapt \senet to be used
within a browser extension. Thus, we first converted the \senet's model
using the TensorflowJS~\cite{TensorFlowJS} tool set. One important factor in
implementing \seguard is performance, especially in terms of
classification latency. Initially, we translated \senet's VGG-19 based
model, which we found to have the best detection accuracy (see Section~\ref{sec:senet_setup}).
Unfortunately, large models such as VGG-19 tend to have poor performance
during inference on regular devices (e.g., a normal laptop) without GPU or other
high-performance hardware. To reduce the latency related to obtaining the
classification decision for a given webpage screenshot, we therefore implemented
a more light-weight version of \senet based on the MobileNet~\cite{MobileNets}
architecture, which provided significant latency reduction with a limited impact
on accuracy. We will present more details on \seguard's evaluation with both
models in Sections~\ref{sec:seguard_setup} and~\ref{sec:seguardeval}.

\begin{comment}
More details on the setup of MobileSENet is provided in section~\ref{sec:seguard_setup}. 

2) Feed the captured screenshot to ''MobileSENetModel'' model that classifies the image as either SE attack or Benign, 3) Display the result of classification to the user via suitable UI based on the result. More notably, SEGuard could also be integrated into the browser as a built-in feature, similar to Chrome's visual based built-in framework to detect Phishing attacks~\cite{google_phishing}.

On that note, we needed to ensure that SEGuard is capable of detecting if a page is SE(or not) at an acceptable speed without much delay while not interfering with user's browsing. Therefore, we build a lighter version of SENet model named ''MobileSENet'', that can be converted and integrated into the extension for faster processing of images and SE detection.  
\end{comment}

\section{Experimental Setup}
In this section, we discuss how we setup the \seshield framework's modules to enable our evaluation.

\subsection{\secrawler Setup}
Our goal is to supply \secrawler's browser farm with seed URLs that are likely to lead to SE attack web pages. To this end, we use the following setup.

\begin{comment}
Our intent with the added improvements in SECrawler is to reach more SE attack content, by visiting a given seed URL and processing its elements within a limited amount of time. In order to achieve that, it is important that we supply SECrawler with seed URLs that yield a higher number of leads to SE attack websites. We discuss the targeted approach made for collecting Seed URLs, SE websites and benign websites as follows
\end{comment}

{\bf Environment Setup}: We use Docker to run many parallel instances of
\secrawler's instrumented browser. Each instance represents an isolated and
clean environment that does not retain any session information related
to crawling previous seed URLs. We deploy our crawlers on a server with 24 CPU
cores running Ubuntu Linux, where we run 15 crawler instances at a time. For any
given seed URL, we visit it using multiple instances of our instrumented
browser, each configured so to emulate one of 10 different browser/device
combinations, including different smartphones, tablets, laptops, etc., and
browsers that render pages with different viewport sizes (the browser/OS
combinations we use include Firefox on Windows, Safari on Mac, Edge on Windows,
Chrome on Windows, Chrome on Linux, Chrome on Mac, Chrome on an Android Phone,
Safari on iPhone, Safari on iPad, and Chrome on different Android tablets).
Additionally, to make our crawling as stealthy as possible and make the
anti-crawler mechanisms implemented by some ad networks more difficult, we used
Puppeteer's ``stealth'' libraries~\cite{extra}.

{\bf SE Seed URLs}: To collect seed URLs that may provide a high-yield of SE attacks, we follow an approach similar to previous studies~\cite{Vadrevu_IMC19,WebPushAds} and collect URLs of websites that are known to host ads from low-tier ad networks, as they have been observed to often lead to SE attack pages. We compiled a list of such ad networks and leveraged PublicWWW~\cite{publicwww}, a reverse code search engine, to find URLs of websites that contained code-related keywords associated with each ad network. Overall, we collected a total of 24,979 seed URLs associated with 10 low-tier ad networks (listed in Table~\ref{tab:ad_networks}, in Appendix).

% In addition, we apply a filtering step on the seed URLs and  preprocess the data to improve our efficiency. 

%  {\em 181,210} seed URLs associated with {\em 21} low tier ad networks.

%  A distribution of the URLs across various low tier ad networks is shown in Table~\ref{tab:ad_networks}. 

\begin{comment} %%% Too many details here %%%
{\bf Filtering seed URLs}: Owing to an extensive and time-consuming crawling
process, we apply a filtering step to eliminate seed URLs that are unlikely to
lead to any SE attacks. As such, we first use a more lightweight version of our
crawler to visit each of the 24,979 seed URLs. The lightweight crawler avoids
time-consuming crawling steps, such as capturing screenshots, but collects
information such as all visited URLs. Next, we only select those seed URLs that
led our crawler towards at least one navigation on a third-party domain. This
filtering step resulted in a reduced set of 6,602 URLs that lead to third-party
domain navigations, on which we ran the full version of \secrawler.
\end{comment}

{\bf SE Attacks Collection}: By visiting the seed URLs, our crawlers collected web pages related to 55,539 distinct URLs and took 650,255 distinct screenshots. Notice that \secrawler takes multiple screenshots per pages, for instance right before the crawler interacts with the page (e.g., click on an element), for every newly opened tabs resulting from the interaction, and in case of changes in the HTML of the current visited page that result from an interaction. In addition, to increase dataset diversity, we also added 28,923
screenshots provided to us by the authors of~\cite{Vadrevu_IMC19}, for a total of 679,178 images. We further process the screenshots to remove duplicates, and cluster images based on their resolution and by using an image similarity metric based on perceptual hashing, similar to previous work~\cite{Vadrevu_IMC19, TRIDENT}. After filtering, we were left with 8,107 images in 318 clusters that needed to be labeled.
To label these image clusters, we relied on a systematic manual labeling process with three expert labelers per image (as described in Section~\ref{sec:details})
% {\em \hyperref[labeling1]{Systematic labeling process}} 
to identify SE attack campaigns with high-confidence ground truth labels. As a result, we found 258 clusters of SE attack pages. Overall, these clusters consisted of 7,484 images related to 30 different screen resolutions. By further (manually) grouping the 258 image clusters related to different resolutions into meta-clusters based on attack type similarity, we obtained 74 different SE attack {\em campaigns}. This new dataset is summarized in Table~\ref{Tab:composition}, where we also contrast ours with the dataset provided by~\cite{Vadrevu_IMC19}. Table~\ref{Tab:imres} (in Appendix) reports a more detailed dataset composition breakdown, wherease Table~\ref{Tab:atcat} summarizes the number of SE campaigns and different screen resolutions per each attack category (e.g., notification stealing, scareware, etc.).

% \usepackage{booktabs}
\begin{table}[h]
    \captionof{table}{Overview of SE attacks dataset.}
    \centering %this
    \begin{adjustbox}{width=\columnwidth} %this
    \small %this
    \begin{tabular}{@{}ccccc@{}}
    \toprule
    \textbf{Dataset}     & \textbf{Campaigns} & \textbf{Screen Resolutions} & \textbf{Screenshots} \\ \midrule
    \text{New data}                   & 30                          & 28                   & 2,490                 \\
    \text{\cite{Vadrevu_IMC19}}                       & 44                          & 3                    & 4,994                 \\  \bottomrule
    \text{Total}                       & 74                          & 30 (1 common res.)                    & 7,484                 \\ \bottomrule
    \end{tabular}
    \end{adjustbox} %this
    \label{Tab:composition}
\end{table}

    \begin{table}[h]
        \captionof{table}{SE attack categories represented in our dataset and number of campaigns and different screen resolutions of web page screenshots we collected.}
        \centering %this
        \begin{adjustbox}{width=\columnwidth} %this
        \small %this
        \begin{tabular}{@{}lcc@{}}
        \toprule
        \textbf{Attack Category}       & \textbf{\# of Resolutions} & \textbf{\# of Campaigns} \\ \midrule
        \text{Fake Software Download}      & 27                         & 29                       \\
        \text{Notification Stealing}  & 26                         & 7                        \\
        \text{Service Sign-up Scam}                & 24                         & 20                       \\
        \text{Scareware}              & 11                         & 9                        \\
        \text{Fake Lottery/Sweepstakes}           & 4                          & 6                        \\
        \text{Technical Support Scam} & 2                          & 3                        \\ \bottomrule
        \end{tabular}
    \end{adjustbox} %this
    \label{Tab:atcat}
        \end{table}

{\bf Benign Webpage Collection}:
To collect examples of benign (i.e., non-SE) web pages, we rely on Tranco~\cite{tranco} and select seed URLs for domains with {rank$<=$5,000}. Overall, we collected 396,255 distinct screenshots by navigating through multiple pages on popular websites using different (emulated) devices and screen resolutions, as for the SE attacks collection.

\begin{comment}
On feeding these URLs to SECrawler, even if an interaction with one of these popular domains lead to a URL with domain of {$rank>5000$}, the crawler stops processing those navigated URLs. As a result, our crawler was able to collect around {\em 396,000} distinct screenshots from {\em 5,401} distinct benign URLs.
\end{comment}

% \karthika{@Irfan Please fill these numbers and double check if they are correct. LEt me know if I have missed any numbers}
% {\bf Image Clustering and Filtering}:

\subsection{\senet Setup}
\label{sec:senet_setup}

As we mentioned in Section~\ref{sec:senet}, after conducting preliminary model selection experiments by comparing five different pretrained models with different architectures, we determine that VGG19 performed the best among them. We therefore used VGG19 pre-trained on ImageNet to build \senet.
To perform all \senet experiments, we used a dedicated Dell PowerEdge R750xa server with 256GB memory and two 48GB NVIDIA A40 GPUs running CUDA 11.8 and TensorFlow v2.8.4.

\subsubsection{Hyperparameter Tuning} Before training the full \senet model, we performed numerous systematic experiments using a reduced dataset of 500 real-world SE attacks and 500 benign training images and 100 benign and 100 SE test images 
% (\roberto{How many images per class?})
to tune the model's hyperparameters so to extract the best performance results on a validation dataset. We used the Adam optimizer with an initial learning rate of $1E-3$, and then tested both lower and higher learning rates. Concurrently, we unfroze either all or a subset of VGG19's {\tt block5} and {\tt block4} convolutional layers. Through these experiments we explored the combination of different learning rate and different unfrozen layers and discovered that setting the learning rate $1E-5$ and unfreezing only the {\tt block5} convolutional layers yielded the best outcomes for our \senet classifier.

Because some of the web page screenshots are quite large (e.g., when taken on a 1920x1080 screen resolution), we reduce input images by using a scaling factor while maintaining the same width/hight ratio. After systematically experimenting with different values of $s$ ranging from 2 to 8, we found that using $s=4$ for desktop and laptop screen resolutions and $s=2$ for mobile device resolutions (i.e., smartphone and tablet) gave us an optimal trade-off between training time and detection accuracy. 

After hyperparameter tuning, we conducted different experiments using the full
dataset (see Table~\ref{Tab:composition}) and our localized federated learning
algorithm described in Section~\ref{sec:senet} (Algorithm~\ref{alg:ALG1}). While
the composition of the training and test dataset changed depending on the experiment (see
details in Section~\ref{sec:expresults}), our models were trained using a consistent setup
across all evaluation scenarios during the training process. First, in each case
we applied data augmentation to the training images and formed a training
dataset containing 15,000 benign images and 15,000 SEA images, with
approximately 500 images per each of the 30 resolutions represented in our
overall dataset. For the localized federated learning algorithm described in
Section~\ref{sec:senet} (Algorithm~\ref{alg:ALG1}), we found that using $CC=5$
clients with training epochs $CE=5$ gave us the best results, while the server's
maximum training epochs was set to $GE=50$, with a batch size of 32. 
During our
experiments, we saved a model checkpoint at every 5 epochs and saved the results for
the best performing model checkpoint on the testing dataset. 

\begin{comment}
When
showing the test results, we calculated the true positive rate at 1 percent
false negative rate and called it the detection rate. Because when we browse the
web, wrongly classifying more than a certain percentage of legit benign websites
would not give a user a good experience. Therefore, three experts that labeled
the dataset in our experiments decided to calculate the detection rate at the 1
percent false negatives and we put that score at the center of our evaluation.
Another reason we consider the detection rate at 1 percent false negative rate
is that the F1,recall, precision and accuracy scores we shared are calculated at
a 0.5 detection threshold and are not optimal. As a result, optimizing the
detection threshold to consider the true positive rate at a maximum false
negative rate yields a better evaluation of the model's detection performance. 
\end{comment}

\subsubsection{Data Augmentation} To augment the training dataset, we performed data augmentations to increase the deep learning model's generalization power. Given the training dataset, for each screen resolution we randomly select images (i.e., screenshots) to which we apply a randomly chosen augmentation algorithm among the ones listed below. 
% This allowed us to increase the size of the dataset to 15.000 benign and 15.000 SE (500 screenshots per resolution). 
Because of the nature of the webpage content, we made sure not to corrupt either the readability of the text content or the orientation of the webpage object, when we chose the augmentation functions. We apply random parameters (e.g., brightness level, crop size and crop windows position, etc.) for each image augmentation, to reduce the possibility of obtaining duplicate images. Every image undergoes the following image augmentations: color inversion, gray scaling, random margin cropping, huge level change, saturation, contrast, brightness, and solarization. The goal io to help the model avoid focusing on specific page artifacts, such as specific colors or the exact page size and size of the page margins, which can in turn help prevent shortcut issues related to shortcut learning~\cite{ShortcutLearning}.

\subsection{\seguard Setup}
\label{sec:seguard_setup}
To build \seguard, we implemented a proof-of-concept Chrome browser extension that embeds our \senet classifier. As mentioned earlier, we initially embedded \senet in the extension by translating the VGG19 model to JavaScript (JS) using the TensorFlowJS framework. However, we quickly realized that the inference time was too large, in the orders of a few seconds, and did not meet our near real-time classification goals. Considering that the SENet model based on VGG19 is quite large, with a parameter count of 144 million, it is not a suitable model for implementation in a browser extension. Therefore, we trained a more lightweight version of \senet based on MobileNetV2~\cite{MobileNets}. To this end, we used MobileNetV2 pre-trained on ImageNet and followed a process similar to what described in Section~\ref{sec:senet}. In terms of detection performance on ImageNet, MobileNetV2 is quite comparable to VGG19, as both models achieved a top-1 accuracy of 71.3\% on the ImageNet validation datasets~\cite{kerasKerasDocumentation}. Because MobileNetV2 expects input images of fixed size, we had to modify its architecture to accommodate the classification of images of arbitrary sizes (similar to what we explained in Section~\ref{sec:senet}). We then fine-tuned the model and translated it using TensorFlowJS to embed it in our proof-of-concept extension. With this approach, we were able to reduce inference time by more than an order of magnitude while maintaining high accuracy. We provide more details on the evaluation of \seguard in Section~\ref{sec:expresults}.

\section{Experimental Results}
\label{sec:expresults}

We evaluated \senet to answer a number of research questions (RQs), as indicated in the following subsections. For instance, we evaluated whether \senet can detect new instances (i.e., new SE attack pages) of previously observed attack campaigns, generalize to previously unseen screen resolutions, detect instances of never-before-seen campaigns, etc. 

Please notice that in all tables reported below, ``DR at 1\% FP'' indicates the detection rate at a 1\% false positive rate, whereas F1, Precision, Recall and Confusion Matrix are computed at the default 0.5 detection threshold.

\begin{comment}
\begin{tcolorbox}
\textbf{R1:}  Can \senet successfully identify {\em new} instances of SE attacks belonging to campaigns that were represented in the training data?
\\\textbf{R2:}  Can SENet successfully identify the same instances of SEA campaigns and benign screenshots when they have completely different resolutions than the ones it was trained on?
\\\textbf{R3:}  Can our SENet be able to detect the examples of SEA campaigns and benign screenshots it has never seen during training?
\\\textbf{R5:}  How SEGuard performs in real-world web browsing?
\end{tcolorbox}
\end{comment}

\subsection{\senet Detection Results}
\label{sec:senetdetection1}

\begin{figure}[!htb]
    \centering\includegraphics[width=0.5\textwidth]{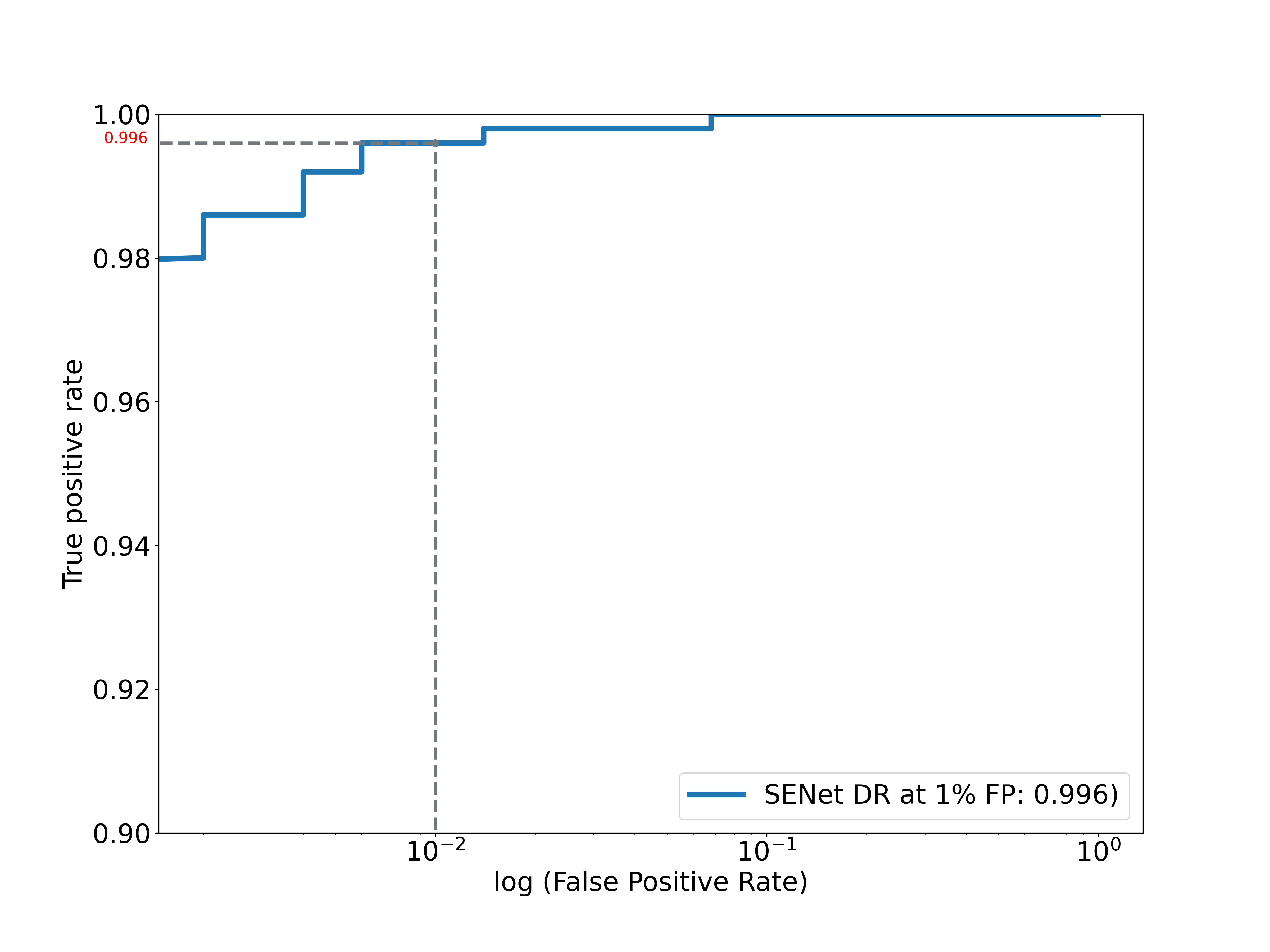}
        \caption{ROC curve for the VGG-19-based \senet model.}
        \label{fig:ROC0}
\end{figure}

\begin{tcolorbox}
  \textbf{RQ1:}  Can \senet accurately identify {\em new} instances of SE attacks belonging to previously observed campaigns?
\end{tcolorbox}
To answer this question, we evaluate the detection capabilities of our \senet model on randomly chosen test data. To this end, we randomly select 500 benign examples 
%irrespective of the URLs come from
and 500 SE attacks, which are kept separate from the training dataset. Notice
that the related web page screenshots were selected across randomly chosen
screen sizes and SE attack campaigns. To compose the training dataset, we
proceeded as indicated in Section~\ref{sec:senet_setup}.
%
\begin{comment}
Because we chose those images randomly for testing, the model would see some of them first time, some of them under different resolutions during training. It is important to acknowledge that within a metacluster, the images possess minor differences in their designs. When we remove certain images from a metacluster for testing purposes, the model does not encounter images with the exact same composition during the testing phase. 
\end{comment}
%
The results are reported in Table~\ref{Tab:table1}, and the (zoomed in) Receiver Operating Characteristic (ROC) curve in Figure \ref{fig:ROC0}.
In this setting, \senet achieves a detection rate of 99.6\% at 1\% false positives. 

\begin{comment}
As a result, we have strong confidence in asserting that when the SENet model is implemented in real-world web browsing applications, it will effectively distinguish new instances of benign webpages and SE attacks with slight changes and some examples with completely different composition and resolution by using only visual cues.
\end{comment}

\begin{table}[h]
  \captionof{table}{SENet's detection results.\label{Tab:table1}}
  \centering %this
  \begin{adjustbox}{width=\columnwidth} %this
  \small %this
  \begin{tabular}{@{}ccccccc@{}}
  \toprule
  \textbf{F1} & \textbf{Recall} & \textbf{Precision} & \textbf{Accuracy} & \textbf{Confusion Matrix}       & \textbf{AUC} & \textbf{DR at 1\% FP} \\ \midrule
  0.994       & 0.992           & 0.996              & 0.994             & TN: 498 FN: 4 FP: 2 TP: 496 & 0.999        & 0.996              \\ \bottomrule
  \end{tabular}
\end{adjustbox} %this
  \end{table}

\subsection{Generalization to Never-Before-Seen Screen Sizes}
\label{sec:sec:senetdetection2}

\begin{tcolorbox}
  \textbf{RQ2:}  Can \senet accurately identify {\em new} instances of SE attacks captured on a screen size never seen during training?
\end{tcolorbox}

\begin{comment}
We wanted to see if our model could successfully generalize to the same examples when they have never-before-seen screen sizes or resolutions. Because, in real-world applications, our model can be deployed in devices with different resolutions and viewports than the 30 resolutions we utilized during the training. With that objective in mind, we performed the following experiments.

\paragraph{\textbf{Mixed Mode Experiments}}
\end{comment}

To answer RQ2, we proceeded as follows. First, we selected 9 different screen resolutions from our dataset. Among these 9 resolutions, 5 are landscape resolutions while the other 4 are portrait resolutions related to mobile devices. Let $r_i$ indicate one of these 9 resolutions. Then, we formed 9 different training datasets, $\{T_1, \dots, T_9\}$, where dataset $T_i$ was formed by selecting images from all resolutions except for $r_i$. The remaining images with resolution $r_i$ were set aside as test dataset $S_i$. We then performed 9 training/test experiment by training on $T_i$ and testing on $S_i$ (i.e., images with the excluded resolution $r_i$).
%
\begin{comment}
When excluding a resolution for testing, we ensured the SEA campaigns and benign images in the excluded resolution were seen during the training under different resolutions. 
\end{comment}
%
Notice that, because \secrawler visits the same URLs using different browser viewport sizes, the same SE attack campaigns can typically be observed under different resolutions. Therefore, $T_i$ and $S_i$ can include common instances of benign pages and SE attacks from the same campaigns, but the training and test images have different sizes. This allows us to focus on the effect of web page screenshots taken on never-before-seen screen sizes, rather than the effects of previously unseen benign pages or SE attack campaigns. Also, to obtain a roughly balanced test dataset, were a particular SE attack campaign does not dominate the others, given resolution $r_i$ we selected at most 10 images from each SE attack campaign represented under that resolution and 1,000 random images from benign pages in the same resolution.

\begin{comment}
Because our crawlers crawled the same URLs using different viewports and user agents, we have the same SEA campaigns under different resolutions, and we made sure to have the same SEA campaigns in the testing set to test only the impact of new resolutions on detection performance. In our dataset, we have 25 landscape resolutions having images with wider widths than height, while the other 5 are portrait resolutions having images with smaller widths than height. Out of the 9 total excluded resolutions for testing, 5 are landscape resolutions, while the other 4 are portrait resolutions. We excluded only 4 portrait malicious resolutions rather than 5, because out of 5 portrait resolutions, one of them contains unique SE attack campaigns than the other ones, and excluding it would be against the nature of that experiment because we only wanted to see the impact on the detection performance when testing images has unique resolutions rather than unique content. Because the images we used during the training have both landscape and portrait resolutions, we called this experiment "Mixed Mode." We got at most 10 images from SE attack campaign and 1000 images per benign resolution from the excluded screen sizes during testing. We chose the resolutions to test the model randomly via a script because we didn't want to bias and chose some examples that may or may not be easy for the model to classify. 
\end{comment}

\begin{figure}[!htb]
        % \centering
        % \begin{center}
            % \vspace{18pt}%
            % \caption@setup{format=plain,justification=centering}
            \includegraphics[width=0.5\textwidth,keepaspectratio]{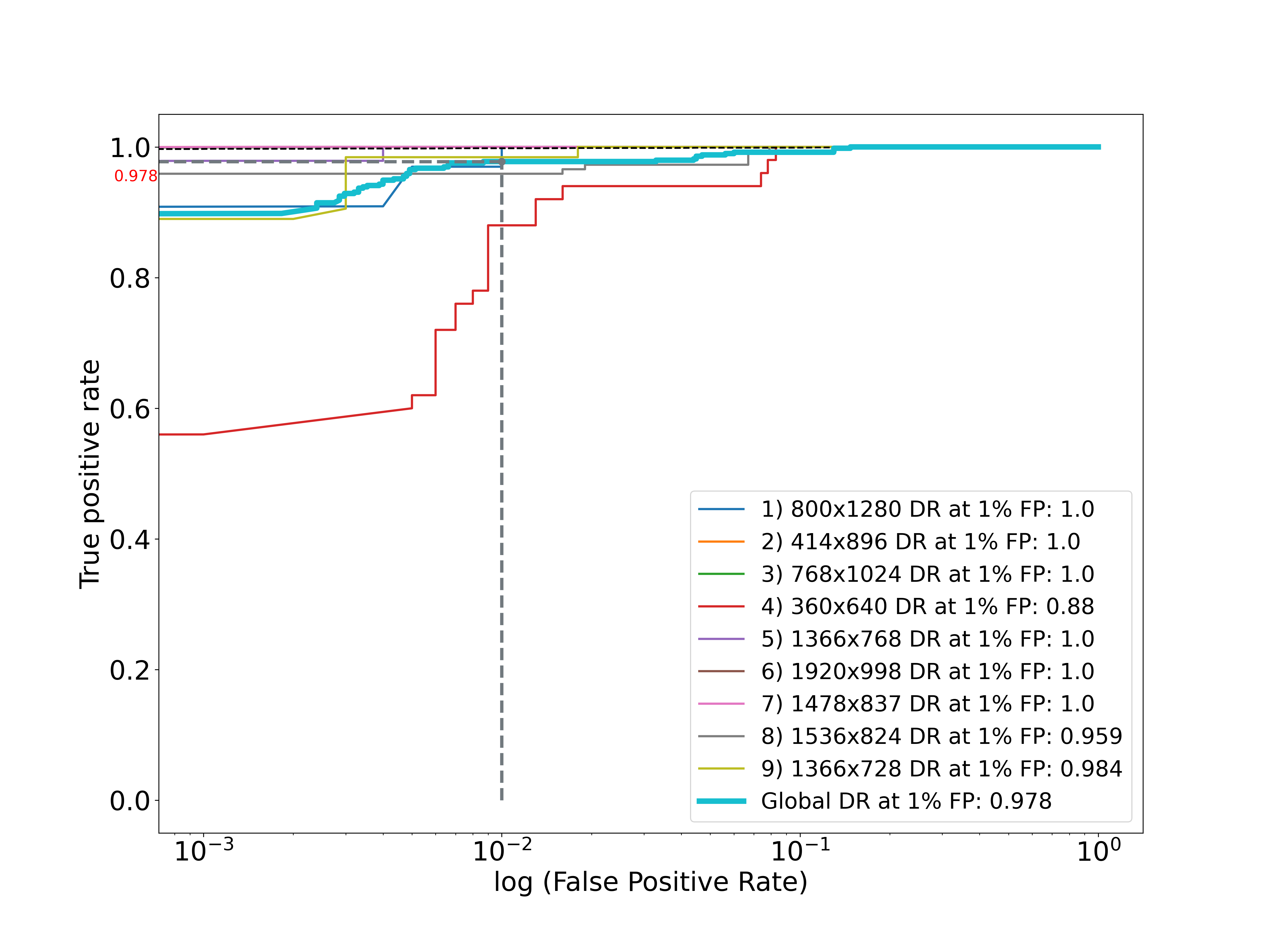}
            
            \caption{ROC curve for the generalization to new screen sizes. Notice that false positives are represented in logarithmic scale, to highlight the classifier's performance at low false positive rates.}
        % \end{center}

        \label{fig:ROC_mix}   
    \end{figure}

The results of this experiment are reported in Table~\ref{Tab:table2} and in Figure~\ref{fig:ROC_mix}. ``Global'' refers to the overall results computed by first combining the classification scores obtained during the 9 training/test experiments into a single scores set. It is notable that \senet can generalize very well to previously unseen resolutions, with detection rates above 95\% at 1\% false positives. A partial exception (88\% detection rate) is represented by {\tt 360x640}, which is typical of older or low-end smartphones. We hypothesize this is due to the particularly small size of this screen resolution, compared to modern devices (notice that the F1, Precision, and Recall values in the table are computed at the default 0.5 detection threshold, rather than the tuned thresholds that would be used in operation to stay below 1\% FPS).

\begin{table}[h]
  \captionof{table}{Results for generalization to new screen sizes. ``DR at 1\% FP'' indicates the detection rate at a 1\% false positive rate. F1, Precision, Recall and Confusion Matrix are instead computed at the default 0.5 detection threshold.} %this
  \centering %this
\begin{adjustbox}{width=\columnwidth} %this
\small %this
  \begin{tabular}{@{}cccccccc@{}}
  \toprule
  \textbf{Test Res.} & \textbf{F1} & \textbf{Recall} & \textbf{Precision} & \textbf{Accuracy} & \textbf{Confusion Matrix}                                                    & \textbf{AUC} & \textbf{DR at 1\% FP} \\ \midrule
  \textbf{800x1280}  & 0.795       & 1.0             & 0.66               & 0.984             & \begin{tabular}[c]{@{}c@{}}TN:983 FN: 0 FP:17 TP:33\end{tabular}       & 0.998        & 1.0                \\
  \textbf{414x896}   & 0.645       & 1.0             & 0.476              & 0.99              & \begin{tabular}[c]{@{}c@{}}TN:989 FN: 0 FP:11 TP: 10\end{tabular}      & 1.0          & 1.0                 \\
  \textbf{768x1024}  & 0.565       & 1.0             & 0.394              & 0.98              & \begin{tabular}[c]{@{}c@{}}TN:980 FN: 0 FP:20 TP:13\end{tabular}       & 0.997        & 1.0                 \\
  \textbf{360x640}   & 0.825       & 0.94            & 0.734              & 0.981             & \begin{tabular}[c]{@{}c@{}}TN:983 FN: 3 FP:17 TP:47\end{tabular}       & 0.992        & 0.88               \\
  \textbf{1366x768}  & 0.94        & 1.0             & 0.887              & 0.994             & \begin{tabular}[c]{@{}c@{}}TN:994 FN: 0 FP:6 TP:47\end{tabular}        & 0.999        & 1.0               \\
  \textbf{1920x998}  & 0.943       & 1.0             & 0.892              & 0.996             & \begin{tabular}[c]{@{}c@{}}TN:996 FN: 0 FP:4 TP:33\end{tabular}        & 0.999        & 1.0                 \\
  \textbf{1478x837}  & 0.941       & 1.0             & 0.889              & 0.995             & \begin{tabular}[c]{@{}c@{}}TN:753 FN: 0 FP:4 TP:32\end{tabular}        & 0.999        & 1.0               \\
  \textbf{1536x824}  & 0.976       & 0.959           & 0.993              & 0.994             & \begin{tabular}[c]{@{}c@{}}TN:999 FN:6 FP:1 TP:140\end{tabular}        & 0.998        & 0.959               \\
  \textbf{1366x728}  & 0.98        & 0.984           & 0.977              & 0.996             & \begin{tabular}[c]{@{}c@{}}TN:997 FN:2 FP:3 TP:125\end{tabular}        & 0.999        & 0.984               \\
  \textbf{Global}  & 0.911        & 0.978           & 0.853              & 0.99             & \begin{tabular}[c]{@{}c@{}}TN: 8674   FN:11 FP: 83 TP:480\end{tabular} & 0.997        & 0.978               \\ \bottomrule
  \end{tabular}
\end{adjustbox} %this

\label{Tab:table2}
  \end{table}

\subsection{Generalization to Never-Before-Seen SE Attack Campaigns}
\label{sec:senetdetection3}

\begin{tcolorbox}
  \textbf{RQ3:}  Can \senet identify web pages belonging to never-before-seen SE attack campaigns?
\end{tcolorbox}

To test whether our model is able to generalize to detecting SE attack pages related to never-before-seen SE attack campaigns, we performed the following experiments.

\begin{figure}[h]
    \captionsetup{justification=centering}
    \begin{subfigure}[t]{.15\textwidth}
      \centering%
       \frame{\includegraphics[width=\linewidth]{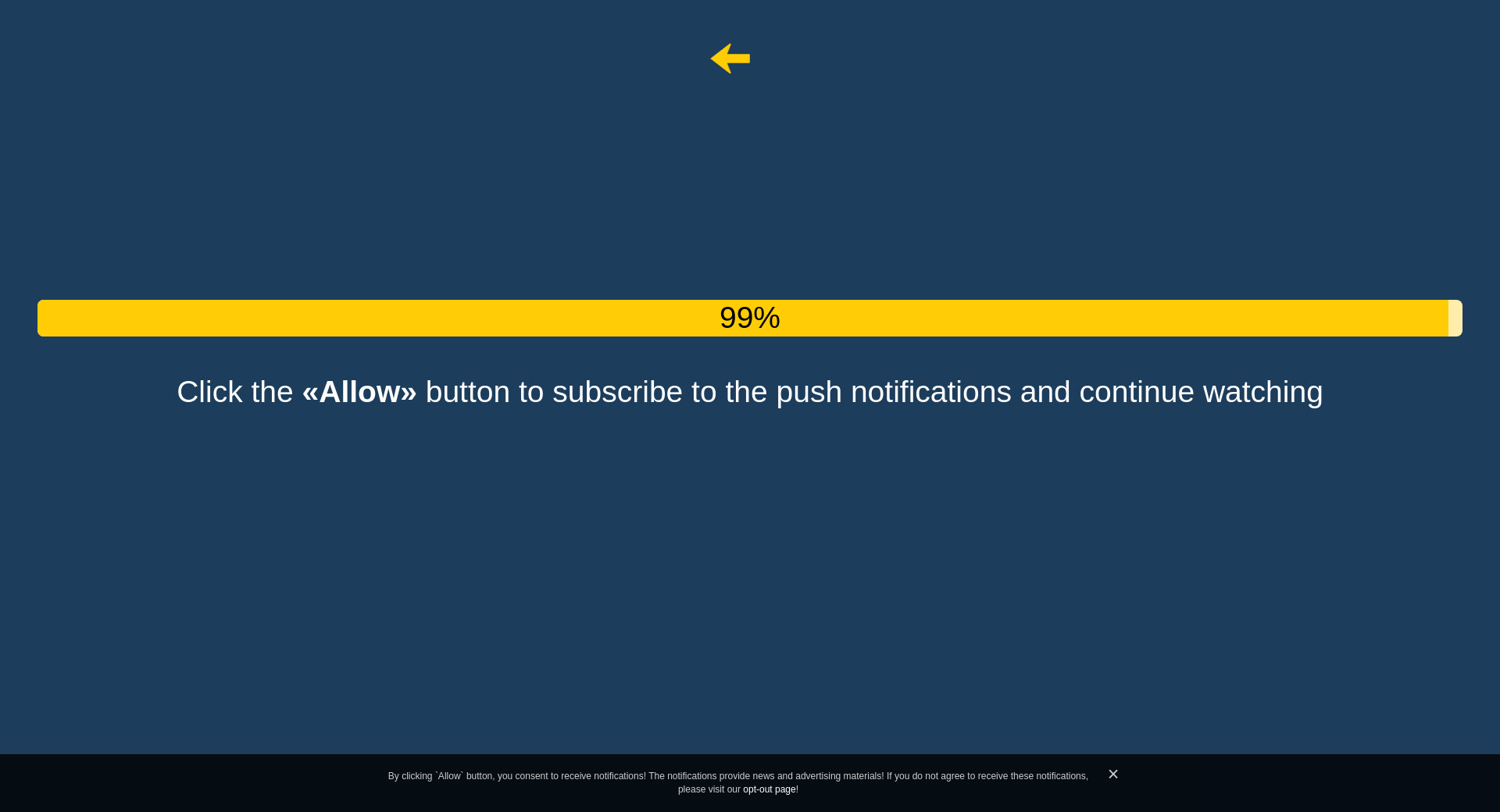}}  
    %   \caption{\protect\raggedright Put your sub-caption here}
      \caption{Notification Stealing I}
      \label{fig:1_SEA}
    \end{subfigure}\hfil % <-- added
    \begin{subfigure}[t]{.15\textwidth}
      \centering%
      \frame{\includegraphics[width=\linewidth]{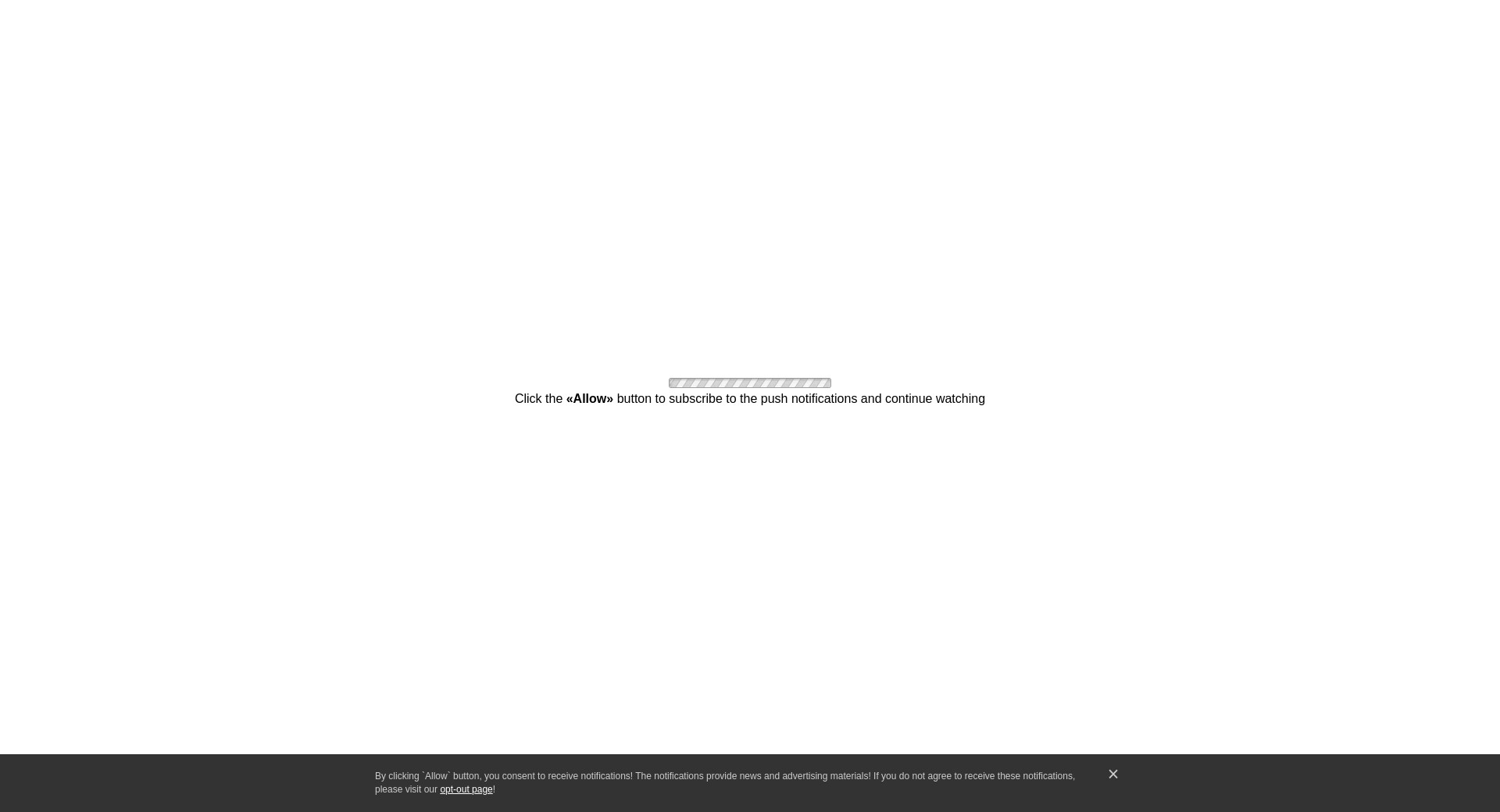}} 
      \caption{Notification Stealing II}
      \label{fig:2_SEA}
    \end{subfigure}\hfil % <-- added
    \begin{subfigure}[t]{.15\textwidth}
        \centering%
        \frame{\includegraphics[width=\linewidth]{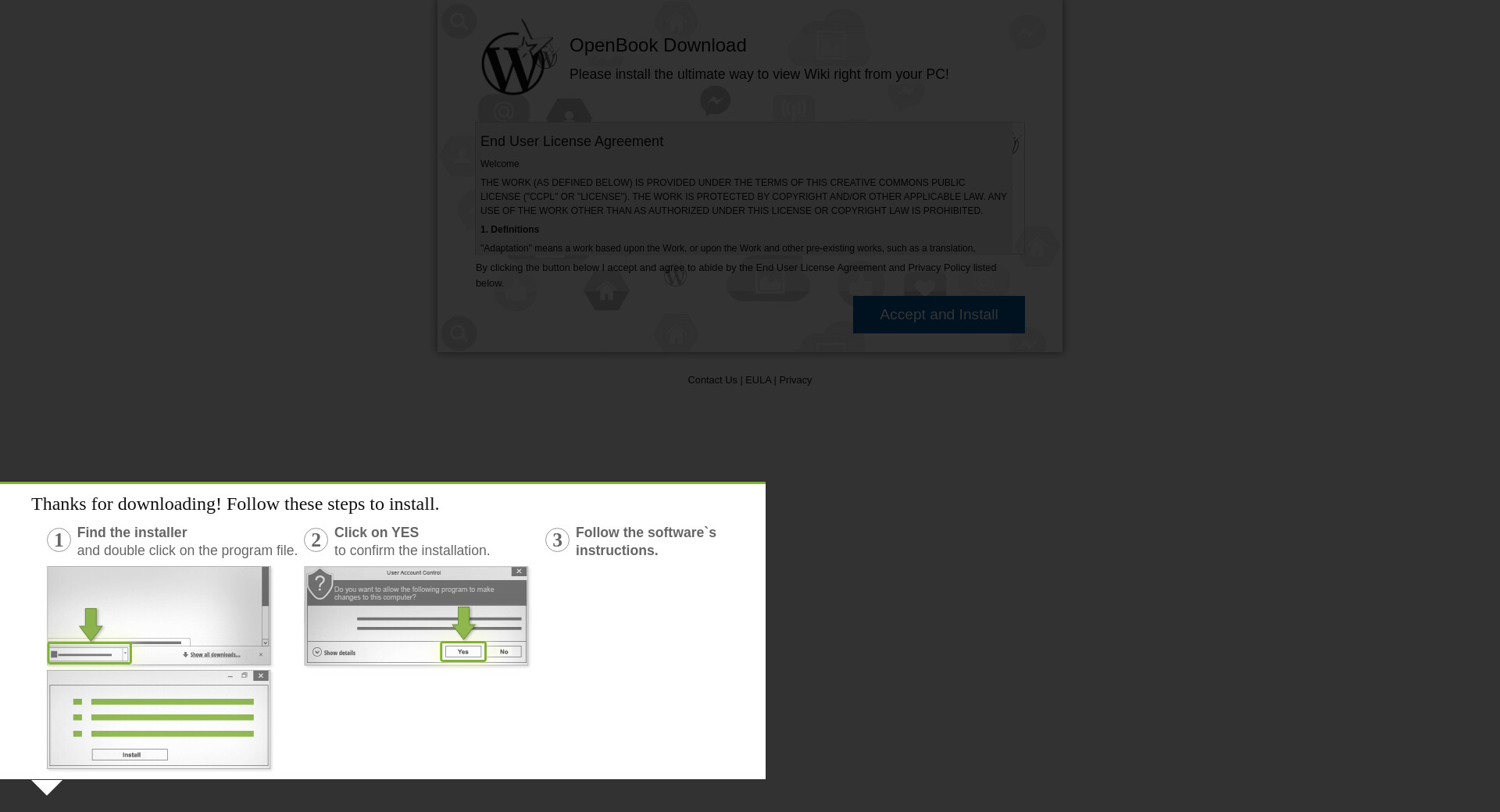}}  
        \caption{Software Download I}
        \label{fig:3_SEA}
      \end{subfigure}\hfil % <-- added
    \newline   
    \begin{subfigure}[t]{.15\textwidth}
      \centering%
      % include third image
      \frame{\includegraphics[width=\linewidth]{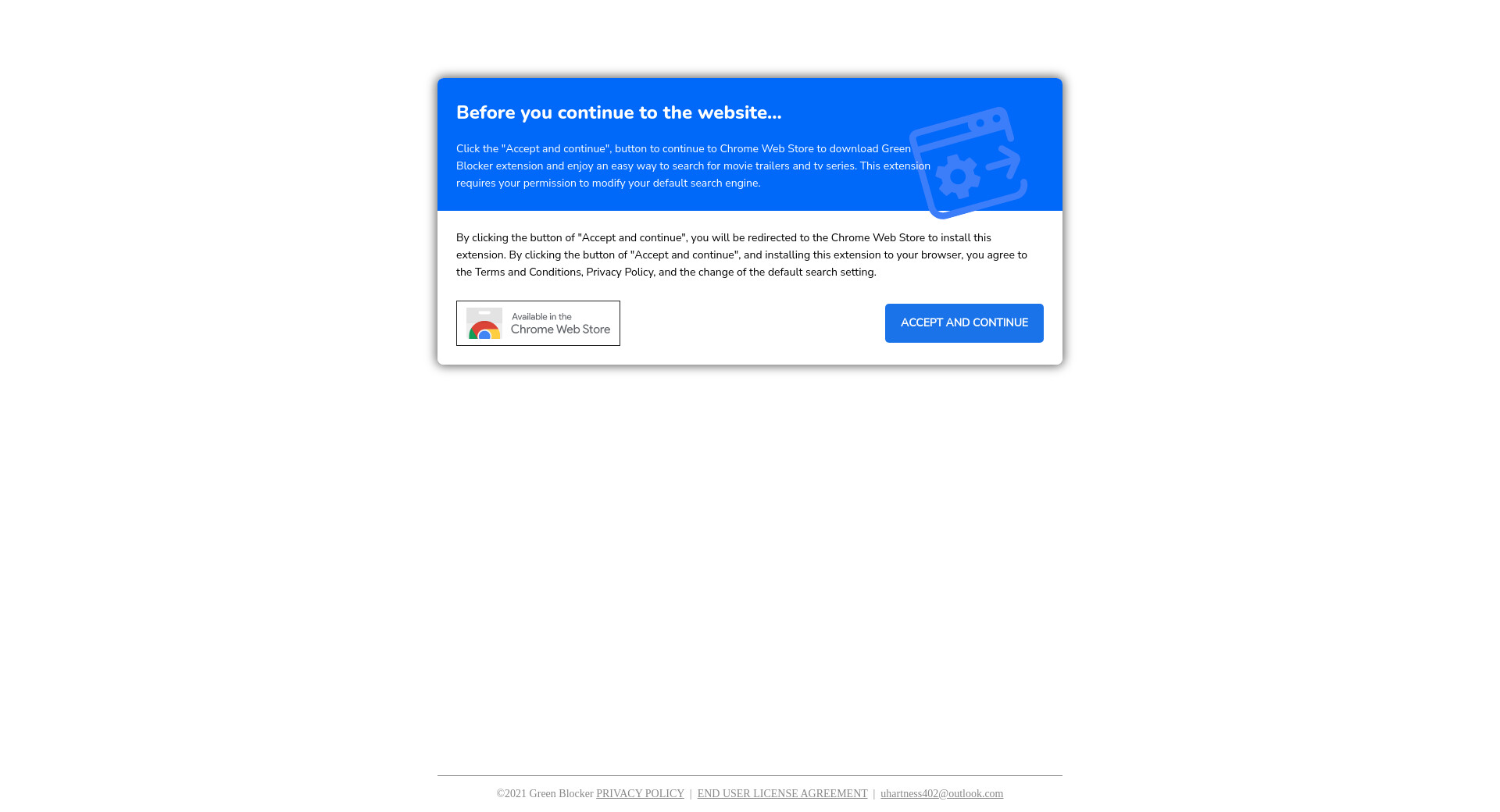}}
      \caption{Software Download II}%
      \label{fig:4_SEA}
    \end{subfigure}\hfil % <-- added
    \begin{subfigure}[t]{.15\textwidth}
      \centering%
      % include fourth image
      \frame{\includegraphics[width=\linewidth]{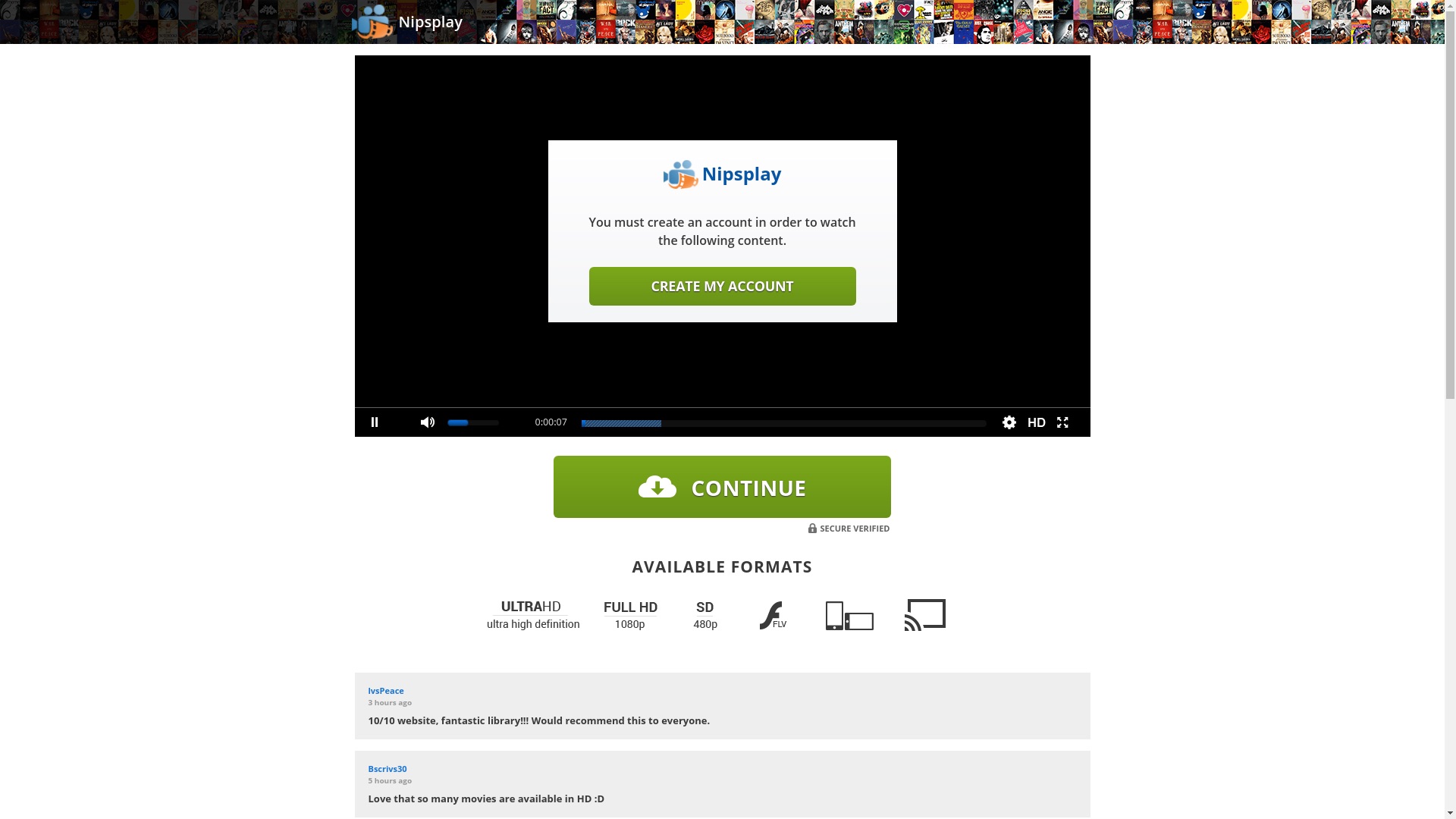}} 
      \caption{Media Streaming}
      \label{fig:5_SEA}
    \end{subfigure}\hfil % <-- added
    \begin{subfigure}[t]{.15\textwidth}
        \centering%
        \frame{\includegraphics[width=\linewidth]{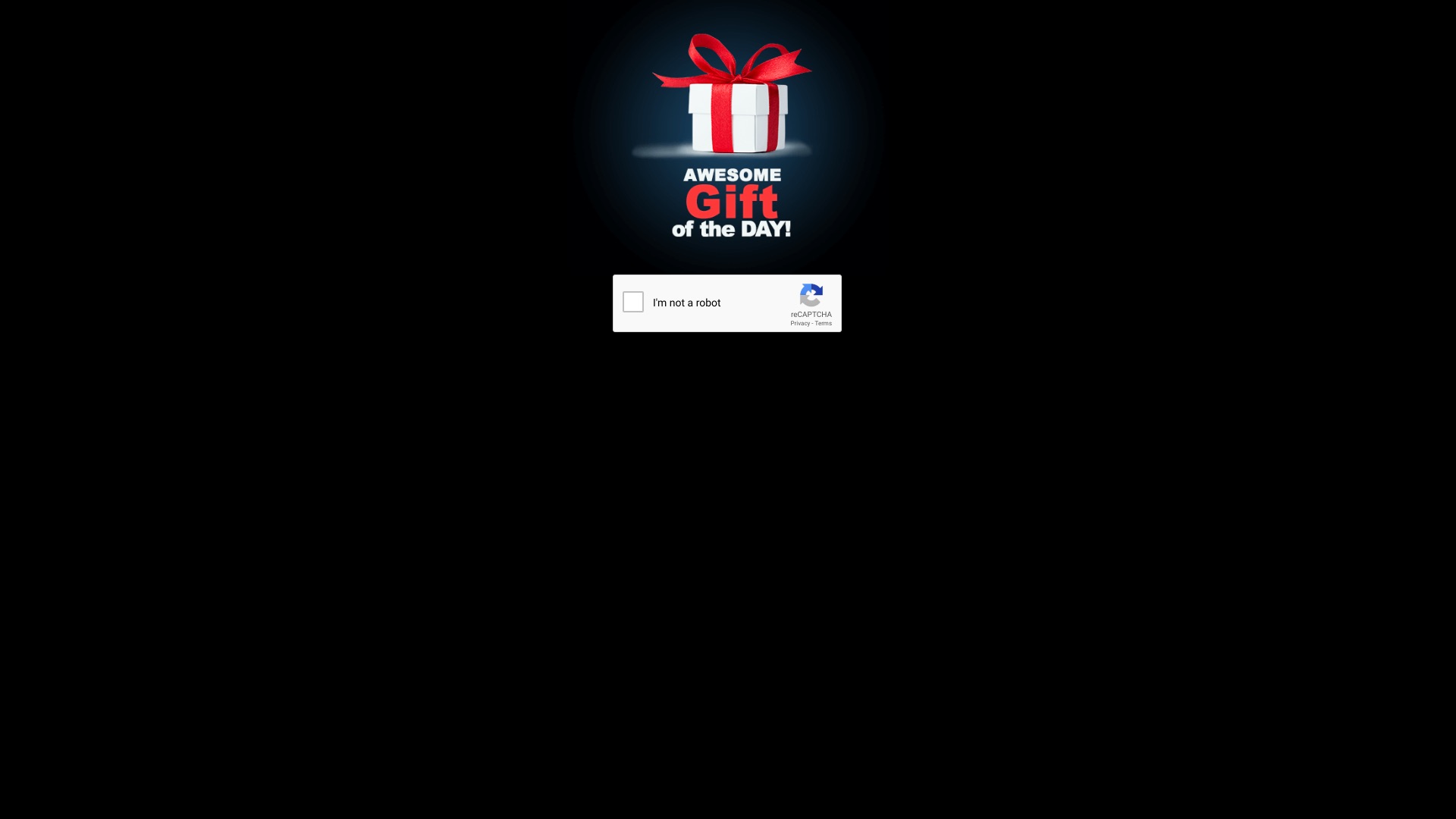}} 
        \caption{Sweepstakes}
        \label{fig:6_SEA}
      \end{subfigure}\hfil % <-- added
      \newline  
      \begin{subfigure}[t]{.11\textwidth}
        \centering%
        \frame{\includegraphics[width=\linewidth]{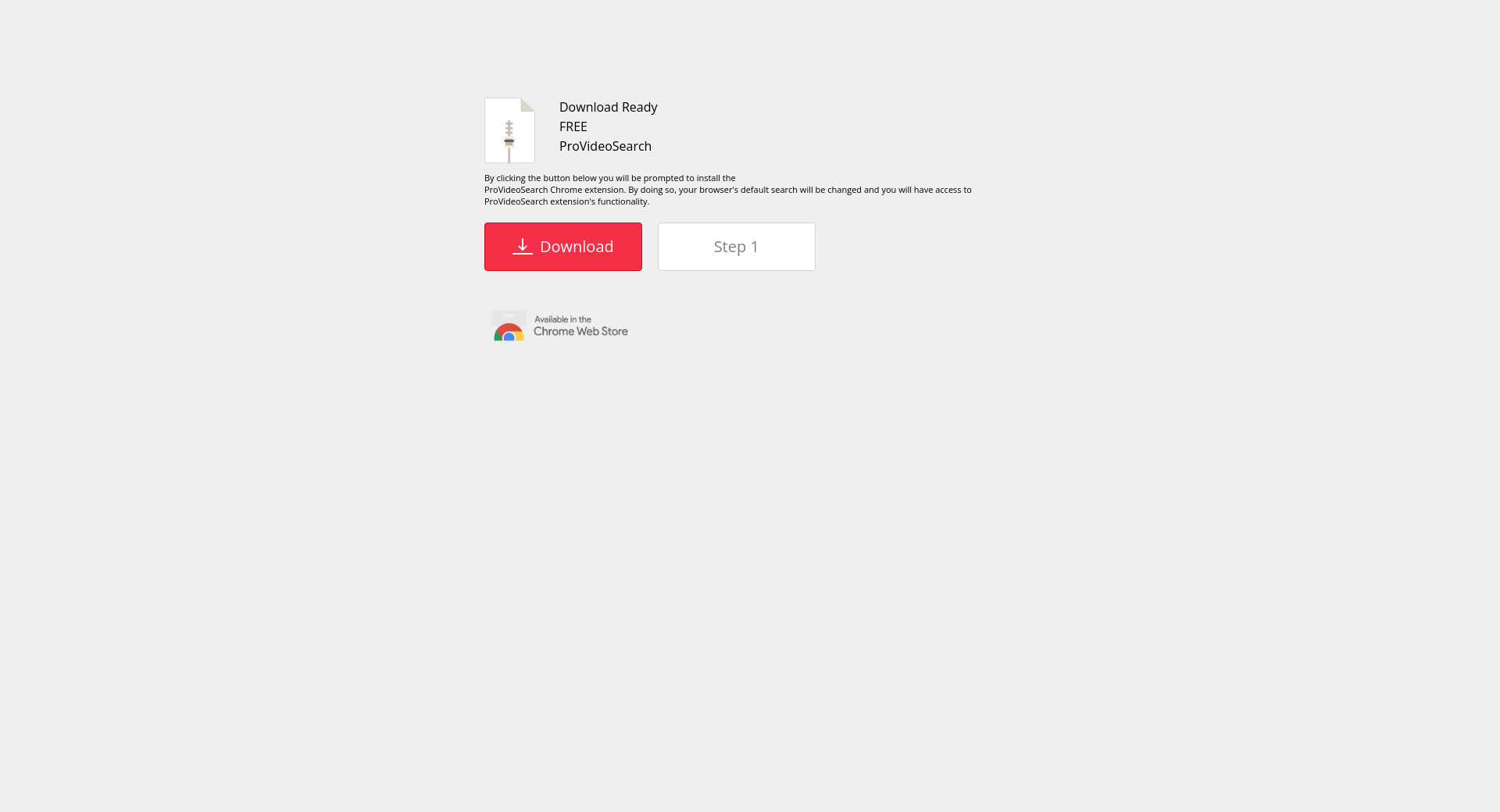}} 
        \caption{Software Download III}
        \label{fig:7_SEA}
      \end{subfigure}\hfil % <-- added
      \begin{subfigure}[t]{.11\textwidth}
        \centering%
        \frame{\includegraphics[width=\linewidth]{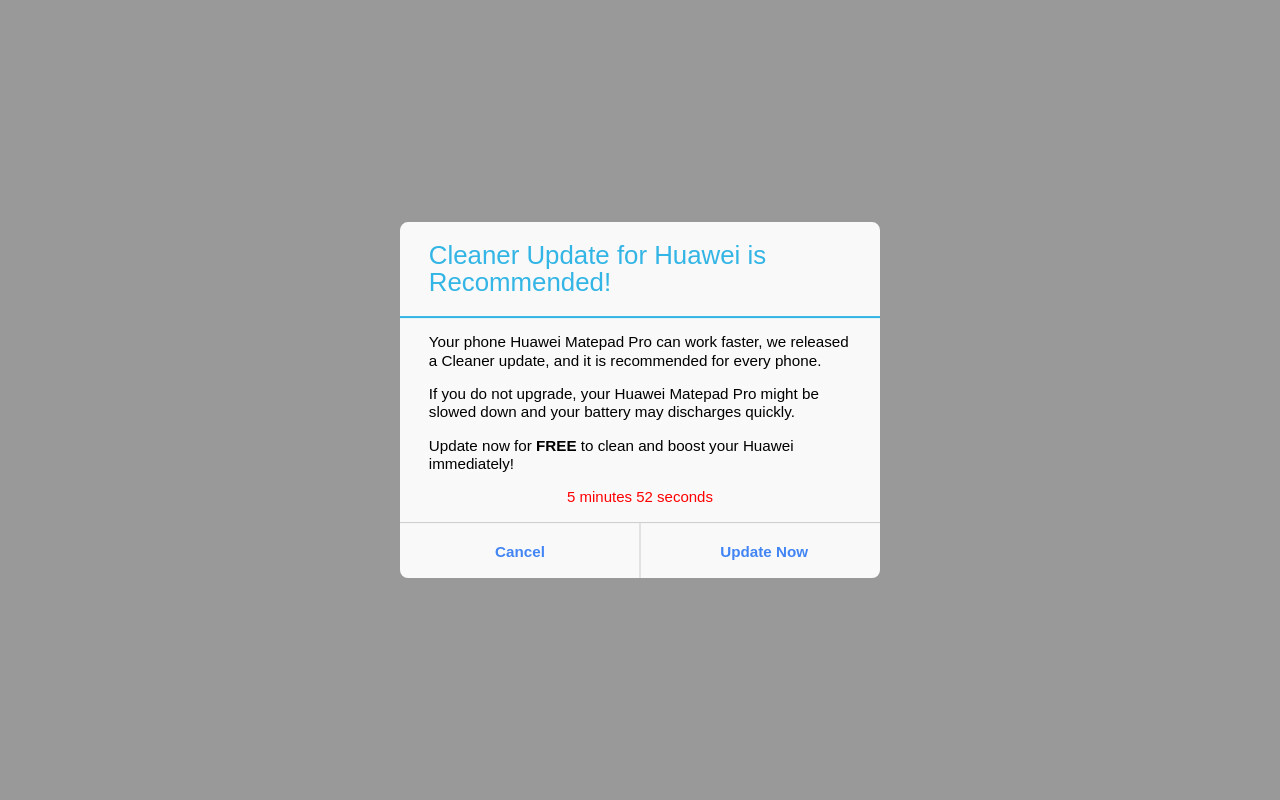}} 
        \caption{Software Download IV}
        \label{fig:8_SEA}
      \end{subfigure}\hfil % <-- added
      \begin{subfigure}[t]{.11\textwidth}
        \centering%
        \frame{\includegraphics[width=\linewidth]{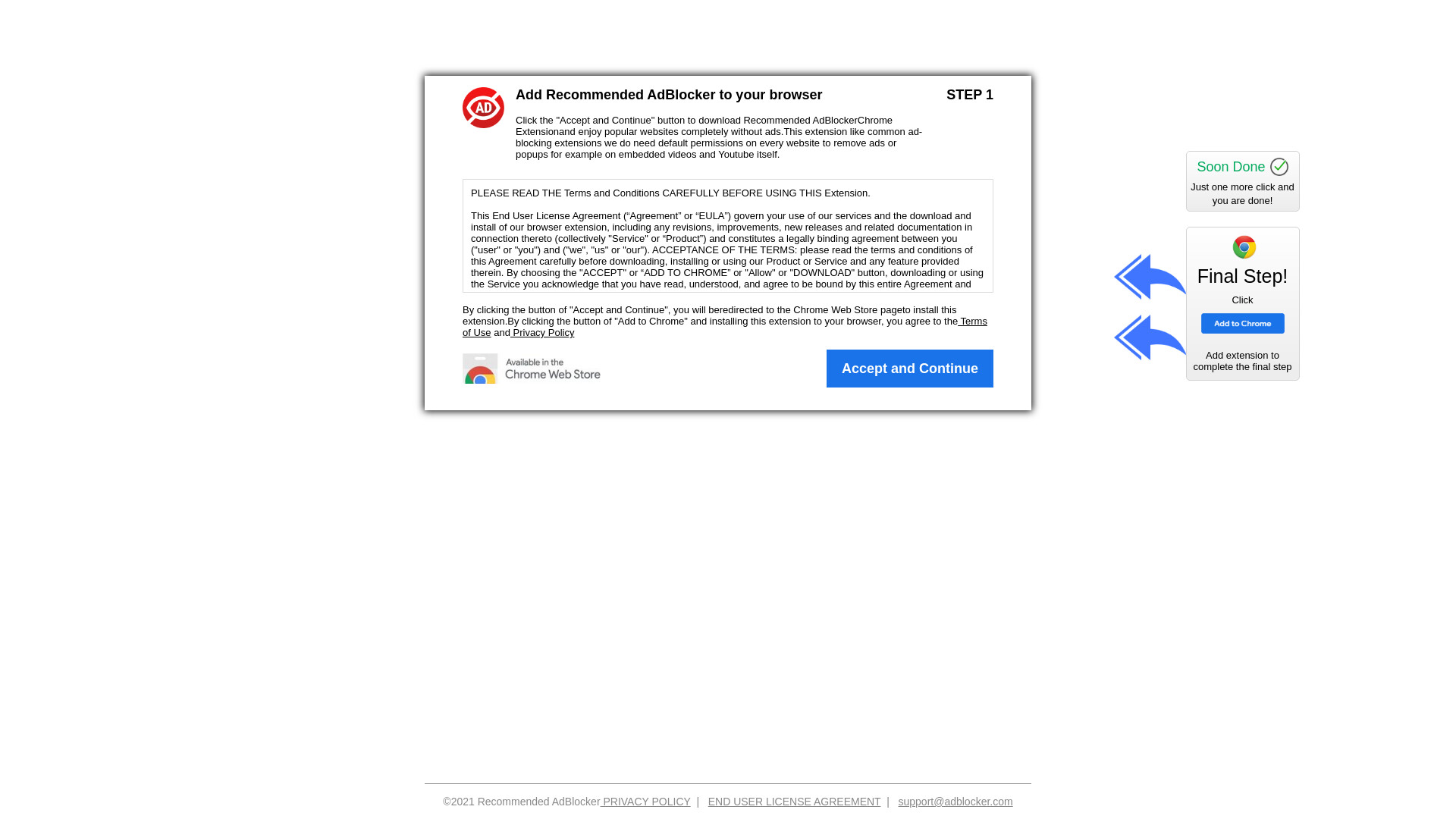}} 
        \caption{Software Download V}
        \label{fig:9_SEA}
      \end{subfigure}\hfil % <-- added
      \begin{subfigure}[t]{.11\textwidth}
        \centering%
        \frame{\includegraphics[width=\linewidth]{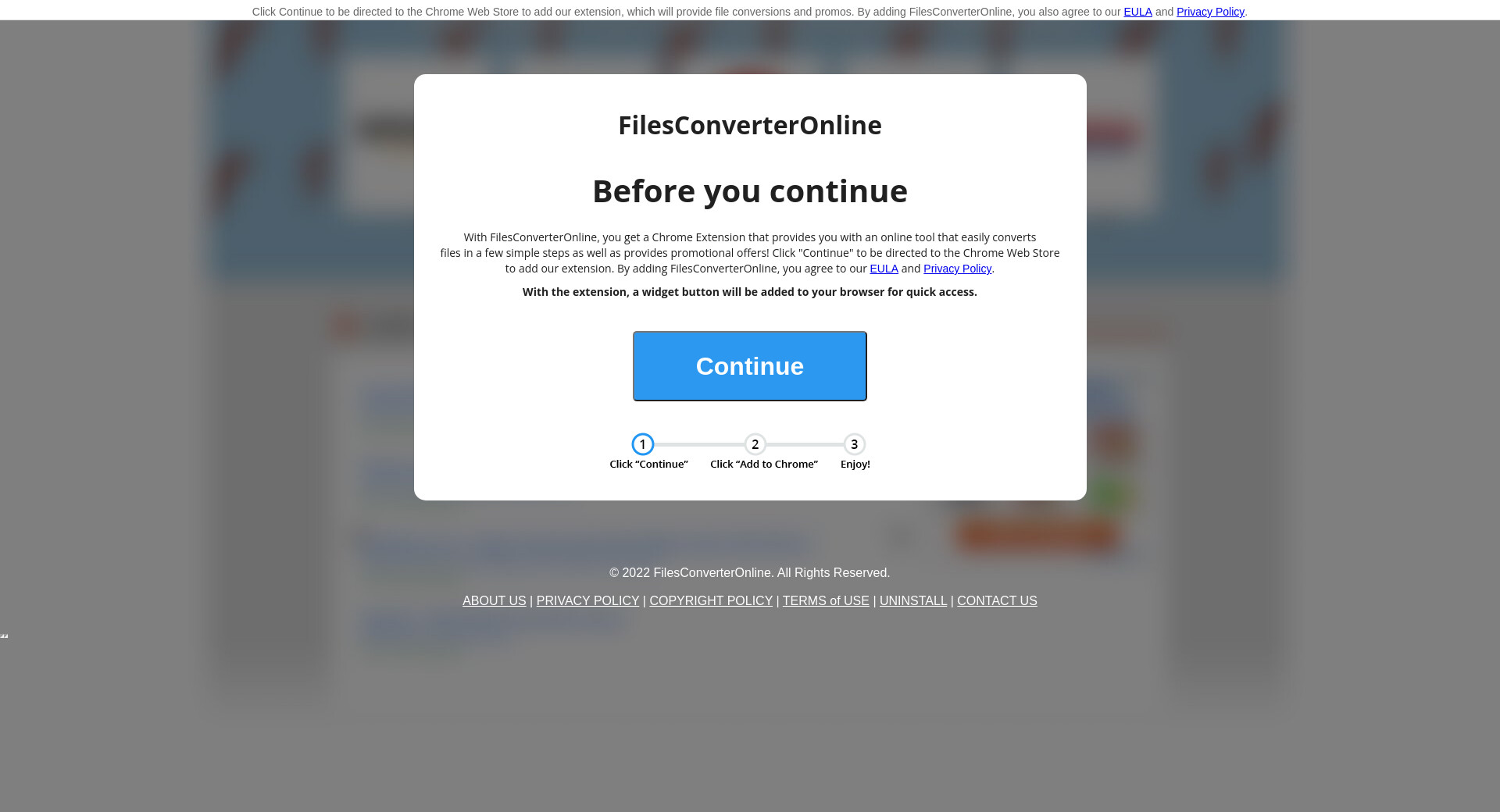}} 
        \caption{Software Download VI}
        \label{fig:10_SEA}
      \end{subfigure}\hfil % <-- added
    \caption{Examples of screenshots from each of the 10 randomly selected test campaigns.}
    \label{fig:MCLUSTERS}
\end{figure}
To setup this experiment, we first randomly selected 10 SEA campaigns, $\{c_1, \dots, c_{10}\}$, to be excluded (in turn) from training and used for testing. Additional, we randomly selected 5000 benign pages from randomly chosen screen resolutions. Example screenshots (one per campaign) for the 10 campaigns $\{c_1, \dots, c_{10}\}$ are shown in Figure \ref{fig:MCLUSTERS}. Notice that while popular SE attack categories, such as Software Download and Notification Stealing are (by chance) represented multiple times, the specific campaigns within those categories are different (e.g., different fake software being distributed and different visual appearance of the attack page).
%
% irrespective of the URLs they come from and from randomly chosen screen sizes.
%
We then trained 10 models, $\{T_1, \dots, T_{10}\}$. For each model, $T_i$, we excluded one SEA campaign, $c_i$, and 500 randomly chosen benign images from training. We then tested the model on the excluded SE campaign and benign images. 
%The amount of SEA testing images depended on how common the attack campaigns were in our seed URLs. 
% \hl{After selecting maximum 10 images per resolution for each of the 10 campaigns, the amount ranges from 15 to 110 SE attack images coming from 3 to 19 different resolutions.} \roberto{I don't understand these numbers. Is it "max 10 images per resolution" for each of the 10 campaigns?}
%
% The selection of SEA campaigns for testing was completely random, , and we can see an example of 10 excluded campaigns in Figure \ref{fig:MCLUSTERS}. 
% 

\begin{figure}[!htb]
  % \centering
  % \begin{center}
      % \vspace{18pt}%
      % \caption@setup{format=plain,justification=centering}
      \includegraphics[width=0.5\textwidth,keepaspectratio]{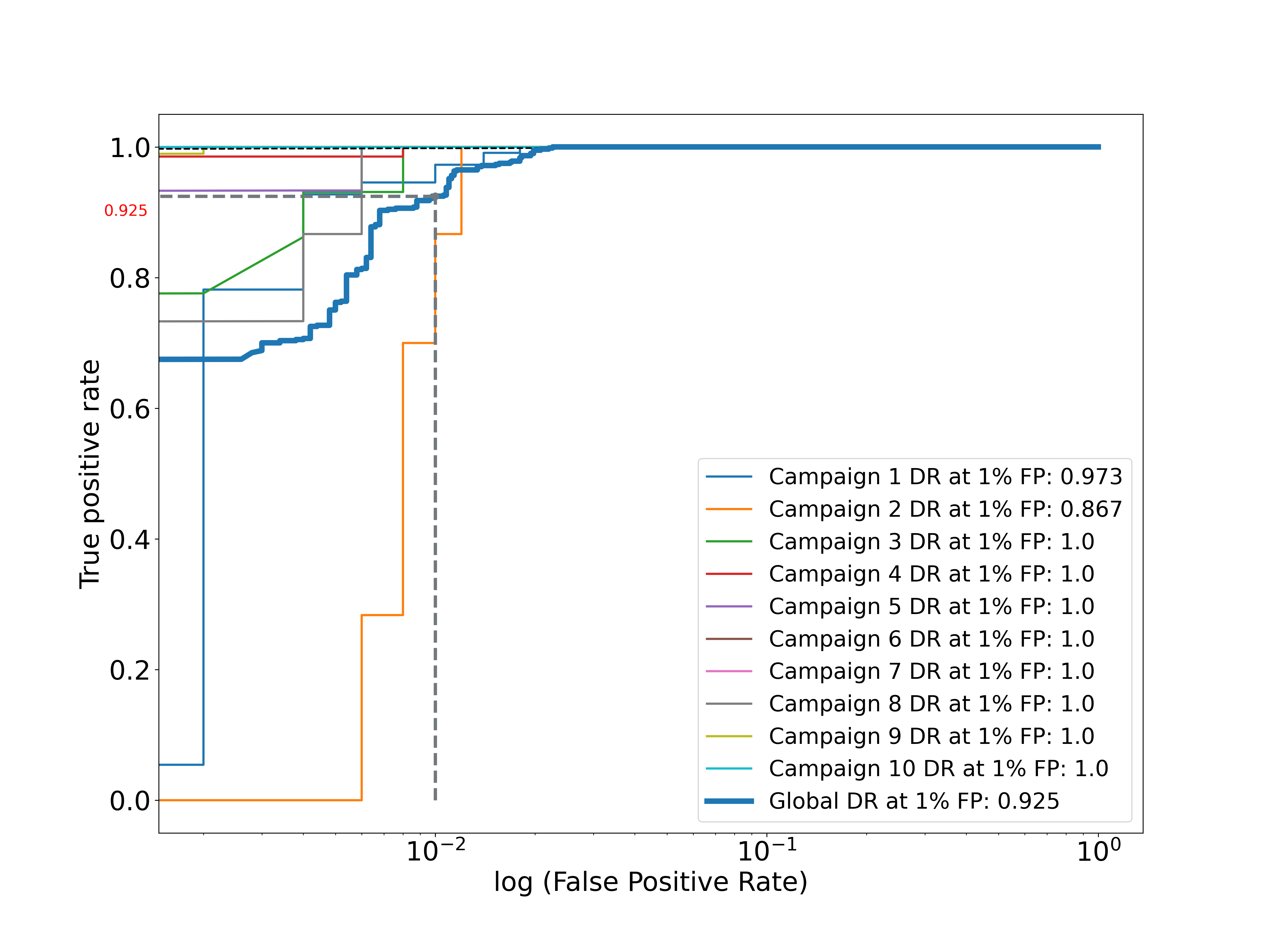}
      
      \caption{ROC Curve for Generalization to Never-Before-Seen SE Campaigns on Single Dataset }
  % \end{center}

  \label{fig:ROC_SINGLESEAALT}    
\end{figure}

The results of this experiment are reported in Table ~\ref{Tab:table41} and in Figure~\ref{fig:ROC_SINGLESEAALT}. We can see that \senet is able to generalize well to never-before-seen attack campaigns, with a detection rate of 92\% at 1\% false positives for all test campaigns but one, namely {\em Campaign 2}. An example screenshot for {\em Campaign 2} is shown in Figure~\ref{fig:2_SEA}. This is a {\em notification stealing} SE attack that shows an almost entirely blank page, which makes it difficult for our classifier to identify significant visual traits that are typical of other campaigns. Nonetheless, more than about 70\% of SE attack page screenshots belonging to this campaign can still be detected by \senet at 1\% false positives.

\begin{table}[h]
  \captionof{table}{Results for the generalization to never-before-seen SE campaigns.\label{Tab:table41}} %this
  \centering %this
\begin{adjustbox}{width=\columnwidth} %this

  \begin{tabular}{@{}cccccccc@{}}
  \toprule
  \textbf{Model name}           & \textbf{F1} & \textbf{Recall} & \textbf{Precision} & \textbf{Accuracy} & \textbf{Confusion Matrix}            & \textbf{AUC} & \textbf{DR at 1\% FP}\\ \midrule
  \textbf{Campaign 1}  & 0.897       & 0.827           & 0.979              & 0.966             & TN: 498 FN: 19 FP: 2 TP: 91    & 0.997        & 0.973               \\
  \textbf{Campaign 2}  & 0.297       & 0.183           & 0.786              & 0.907             & TN: 497 FN: 49 FP: 3 TP: 11    & 0.992        & 0.87              \\
  \textbf{Campaign 3}  & 0.959       & 1.0             & 0.921              & 0.991             & TN: 495 FN: 0  FP: 5 TP: 58     & 0.998        & 1.0               \\
  \textbf{Campaign 4}  & 0.971       & 1.0             & 0.944              & 0.993             & TN: 496 FN: 0  FP: 4 TP: 67     & 1.0          & 1.0               \\
  \textbf{Campaign 5}  & 0.952       & 1.0             & 0.909              & 0.994             & TN: 497 FN: 0  FP: 3 TP: 30     & 0.998        & 1.0              \\
  \textbf{Campaign 6}  & 0.833       & 1.0             & 0.714              & 0.988             & TN: 494 FN: 0  FP: 6 TP: 15     & 0.996        & 1.0               \\
  \textbf{Campaign 7}  & 0.966       & 1.0             & 0.935              & 0.995             & TN: 497 FN: 0  FP: 3 TP: 43     & 0.999        & 1.0                \\
  \textbf{Campaign 8}  & 0.909       & 1.0             & 0.833              & 0.989             & TN: 494 FN: 0  FP: 6 TP: 30     & 0.997        & 1.0              \\
  \textbf{Campaign 9}  & 0.99        & 1.0             & 0.98               & 0.997             & TN: 498 FN: 0  FP: 2 TP: 96     & 0.999        & 1.0               \\
  \textbf{Campaign 10} & 1.0         & 1.0             & 1.0                & 1.0               & TN: 500 FN: 0  FP: 0 TP: 88     & 1.0          & 1.0                 \\
  \textbf{Global}               &0.912        &0.886            &0.94                &0.982              & TN: 4966 FN: 68 FP: 34 TP: 529 & 0.997        & 0.925              \\ \bottomrule
  \end{tabular}
\end{adjustbox} %this
  \end{table}

\subsection{Adversarial Examples Evaluation}

\begin{tcolorbox}
  \textbf{RQ4:}  Can \senet be strengthened against adversarial examples?
\end{tcolorbox}

Because SE attack pages are under the attacker's control, the attacker can attempt to craft attack pages that explicitly attempt to evade \senet by injecting adversarial noise. At the same time, attackers must operate under two concurrent constraints: (i) SE attack pages must maintain their visual attack components, and (ii) the level of injected noise should be limited so not be noticeable to users.

\begin{figure}[h]
  \captionsetup[subfigure]{labelformat=empty}
  \captionsetup{justification=centering}
  \begin{subfigure}[t]{.15\textwidth}
    \centering%
     \frame{\includegraphics[width=\linewidth]{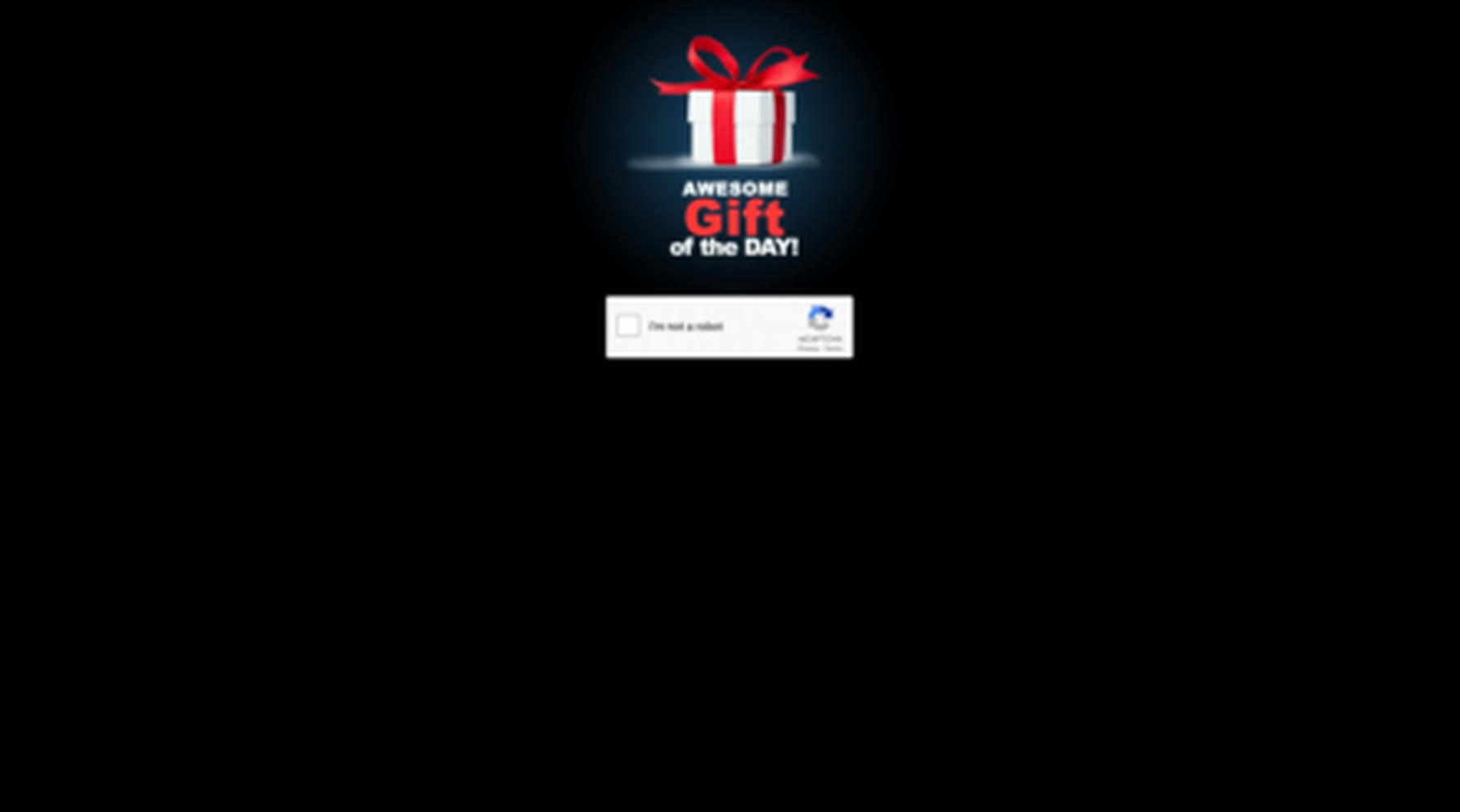}}  
  %   \caption{\protect\raggedright Put your sub-caption here}
    \caption{epsilon=0.3}
    \label{fig:1_PGD}
  \end{subfigure}\hfil % <-- added
  \begin{subfigure}[t]{.15\textwidth}
    \centering%
    \frame{\includegraphics[width=\linewidth]{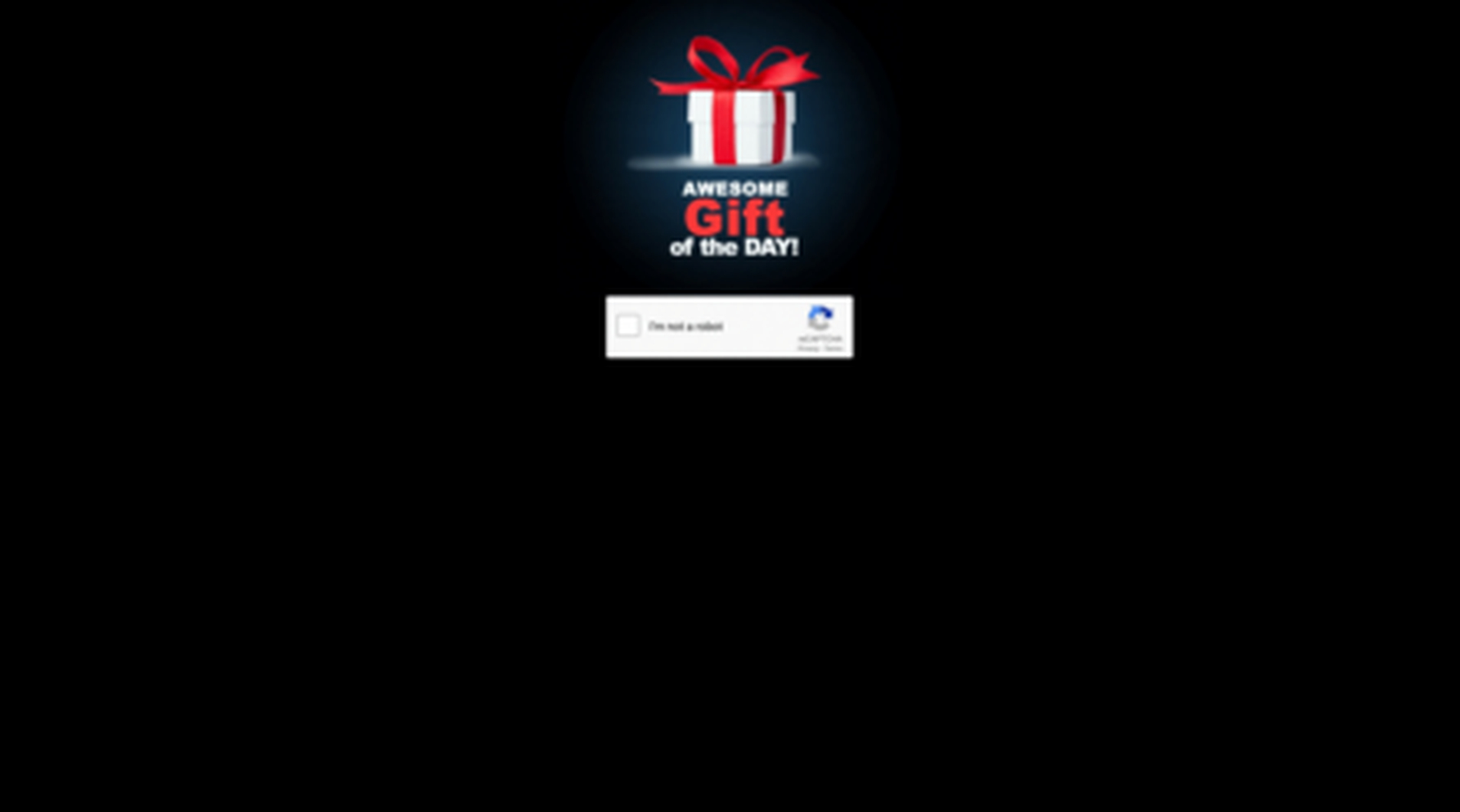}} 
    \caption{epsilon=0.5}
    \label{fig:2_PGD}
  \end{subfigure}\hfil % <-- added
  \begin{subfigure}[t]{.15\textwidth}
      \centering%
      \frame{\includegraphics[width=\linewidth]{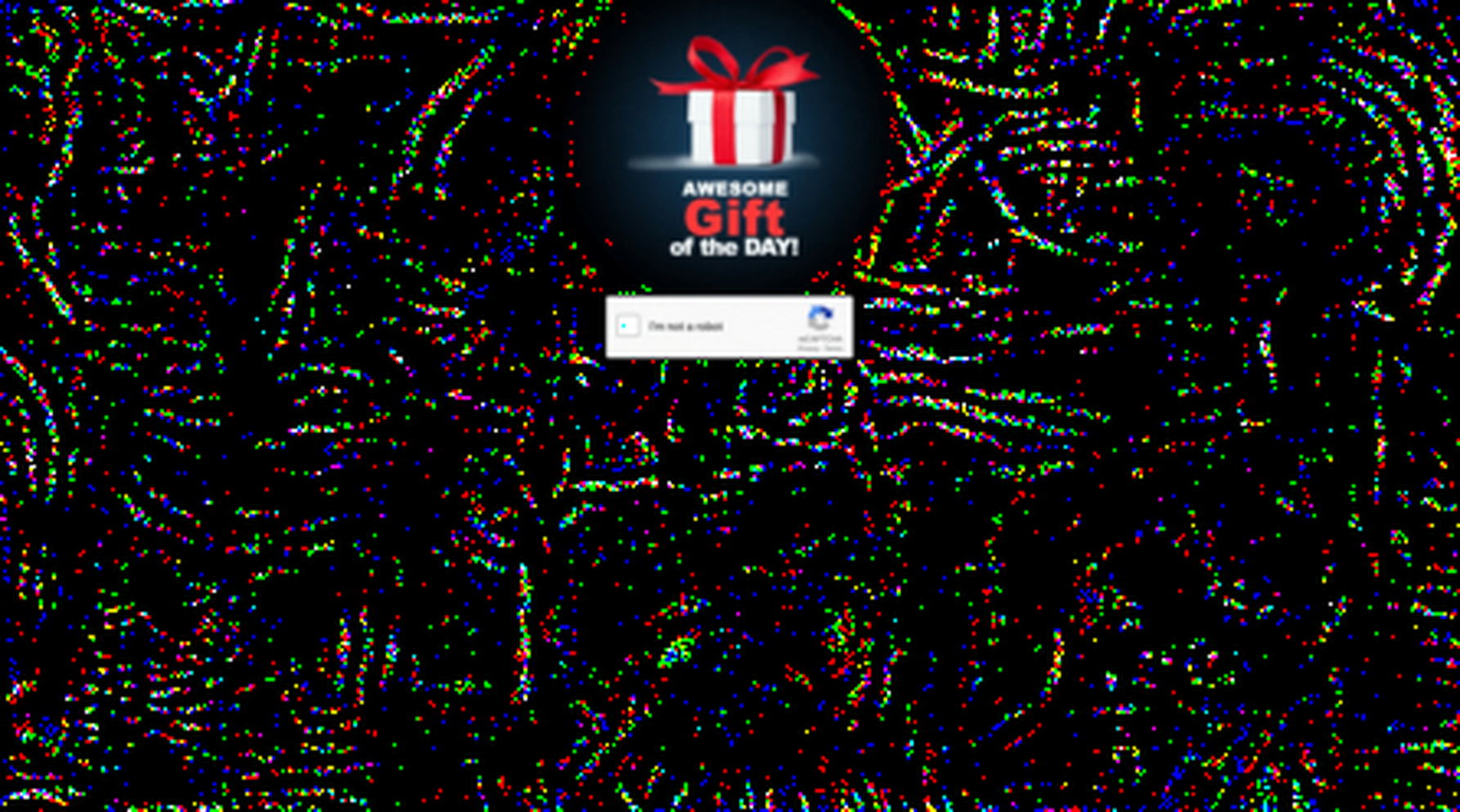}}  
      \caption{epsilon=1}
      \label{fig:3_PGD}
    \end{subfigure}\hfil % <-- added
  \newline   
  \begin{subfigure}[t]{.15\textwidth}
    \centering%
    % include third image
    \frame{\includegraphics[width=\linewidth]{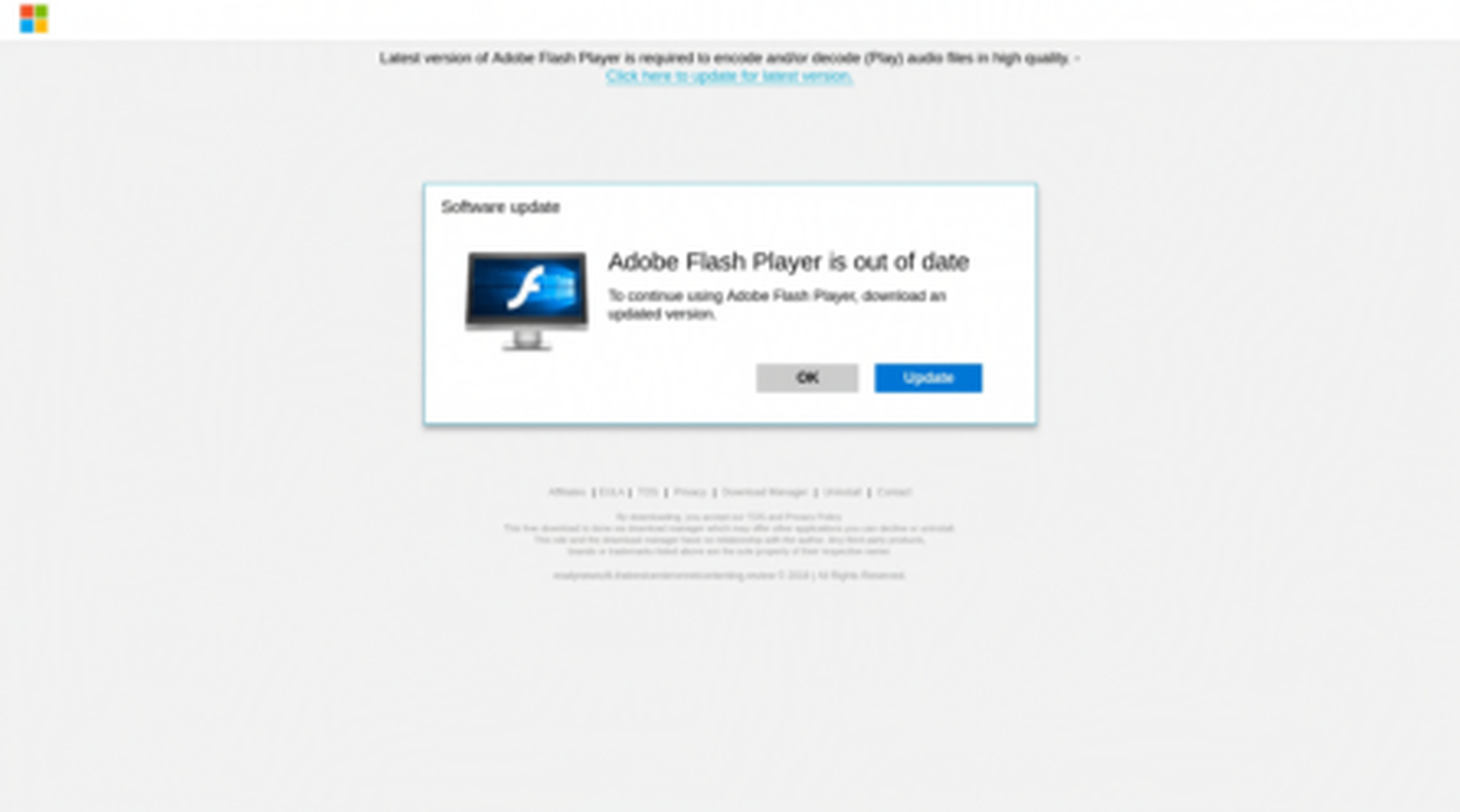}}
    \caption{epsilon=0.3}%
    \label{fig:4_PGD}
  \end{subfigure}\hfil % <-- added
  \begin{subfigure}[t]{.15\textwidth}
    \centering%
    % include fourth image
    \frame{\includegraphics[width=\linewidth]{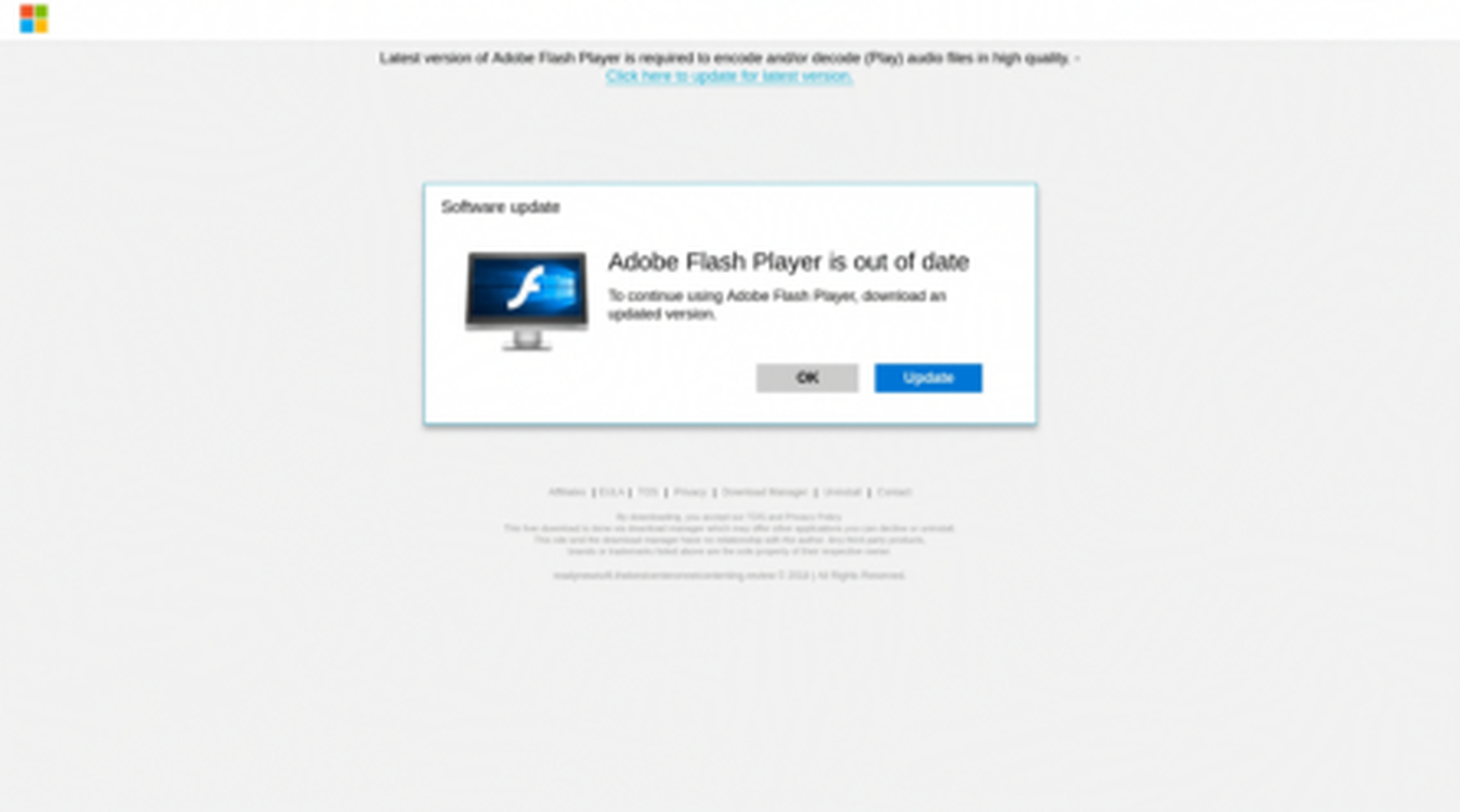}} 
    \caption{epsilon=0.5}
    \label{fig:5_PGD}
  \end{subfigure}\hfil % <-- added
  \begin{subfigure}[t]{.15\textwidth}
      \centering%
      \frame{\includegraphics[width=\linewidth]{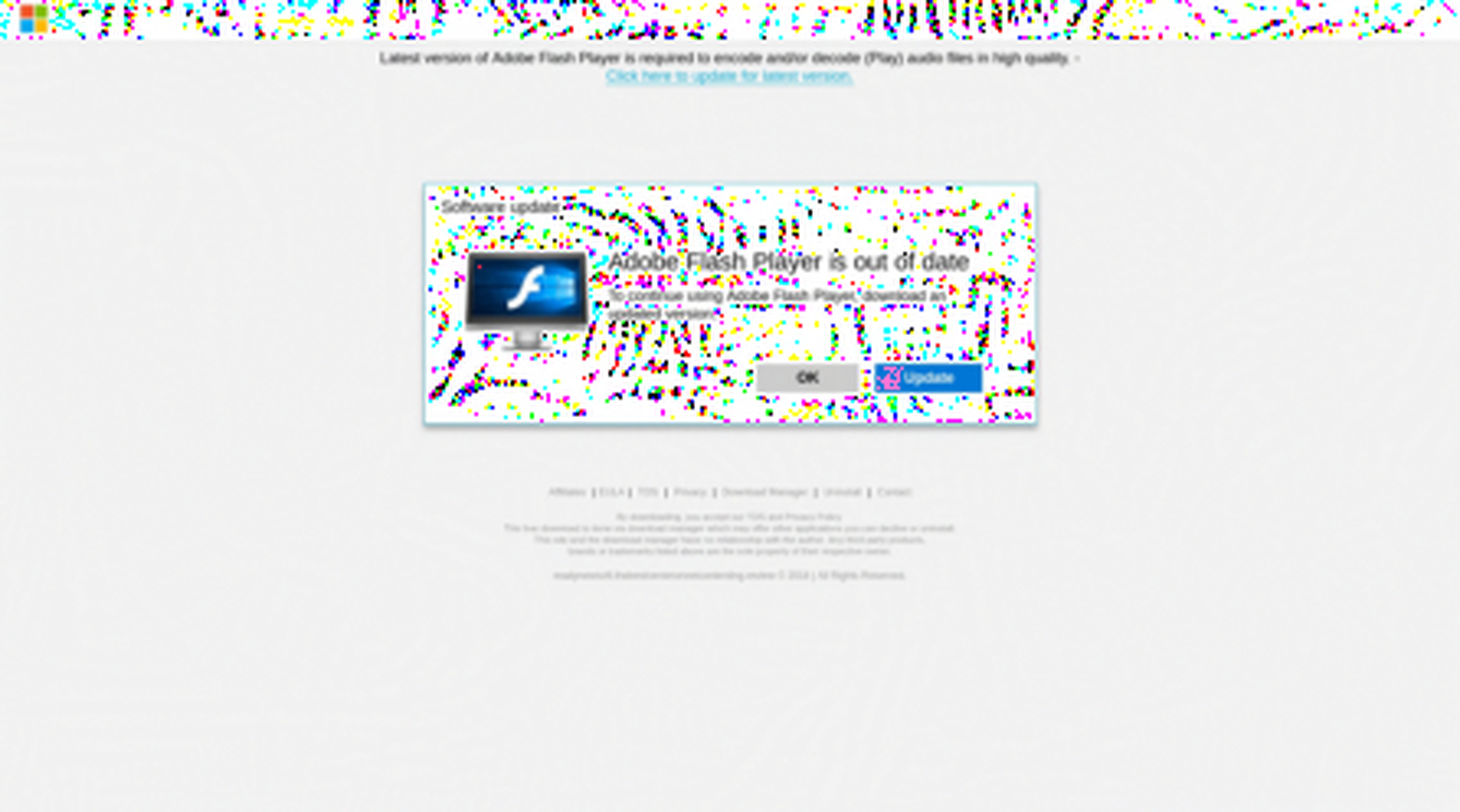}} 
      \caption{epsilon=1}
      \label{fig:6_PGD}
    \end{subfigure}\hfil % <-- added

  \caption{Adversarial Examples Generated with PGD}
  \label{fig:PGD}
\end{figure}

It is well known that adversarial examples forged by adding small perturbations can reduce deep learning models' classification accuracy~\cite{chakraborty2018adversarial} and that adversarial training is an effective approach to make models (at least partially) more robust to them~\cite{ijcai2021bai}. In this section, we first test \senet against white-box adversarial attacks, and then show that adversarial training can significantly increase its robustness to such attacks. To this end, we follow advise provided by highly cited previous work in this area~\cite{madry2017towards} and use projected gradient descent (PGD) to construct adversarial images in a white-box attack setting using the Foolbox library~\cite{rauber2017foolboxnative,rauber2017foolbox}.
%
\begin{comment}
to computes the gradient of the model's loss function concerning the input image and introduces a small perturbation aligned with the gradient's direction. This perturbed input resembles the original but results in misclassification by the model. During our adversarial testing and training we used the Foolbox Python Library~\cite{rauber2017foolboxnative,rauber2017foolbox}.
\end{comment}
%
The level of perturbation is controlled by the $\epsilon$ parameter. The higher the value of $\epsilon$, the more noticeable the perturbation to the human eye. At the same time, larger $\epsilon$ values increase the likelihood of evading a deep learning model.
%
\begin{comment}
Using larger $\epsilon$ values increases the likelihood of deceiving a deep learning model, but it diminishes the overall effectiveness of the attack due to the noticeable nature of the perturbations, rendering the victim less susceptible. 
\end{comment}
%
Therefore, we experimented with increasing values of $\epsilon$, starting from 0.01. As shown in Figure~\ref{fig:PGD}, as $\epsilon$ reaches 1 the noise becomes very noticeable.

\begin{table}[h]
  \captionof{table}{Results for different PGD $\epsilon$ values.\label{Tab:table8}} %this
  \centering %this
\begin{adjustbox}{width=\columnwidth} %this
\small %this
  \begin{tabular}{@{}ccccccc@{}}
  \toprule
  \textbf{$\epsilon$} & \textbf{F1} & \textbf{Recall} & \textbf{Precision} & \textbf{Accuracy} & \textbf{Conf.   Matrix}       & \textbf{DR at 1\% FP} \\ \midrule
  \textbf{NONE}    & 0.994       & 0.992           & 0.996              & 0.9936            & TN: 433 FN: 4   FP: 2 TP: 496 & 0.996                          \\
  \textbf{0.01}    & 0.993       & 0.99            & 0.996              & 0.9925            & TN: 433 FN: 5 FP: 2 TP: 495   & 0.996                          \\
  \textbf{0.1}     & 0.99        & 0.984           & 0.996              & 0.9893            & TN: 433 FN: 8 FP: 2 TP: 492   & 0.988                          \\
  \textbf{0.3}     & 0.8732      & 0.778           & 0.9949             & 0.8791            & TN: 433 FN: 111 FP: 2 TP: 389 & 0.8                            \\
  \textbf{0.5}     & 0.7145      & 0.558           & 0.9929             & 0.7615            & TN: 433 FN: 221 FP: 2 TP: 279 & 0.616                          \\
  \textbf{1}       & 0.4139      & 0.262           & 0.985              & 0.6032            & TN: 433 FN: 369 FP: 2 TP: 131 & 0.27                           \\
  \textbf{2}       & 0.4089      & 0.258           & 0.9847             & 0.6011            & TN: 433 FN: 371 FP: 2 TP: 129 & 0.258                          \\
  \textbf{3}       & 0.4114      & 0.26            & 0.9848             & 0.6021            & TN: 433 FN: 370 FP: 2 TP: 130 & 0.26                           \\
  \textbf{4}       & 0.4038      & 0.254           & 0.9845             & 0.5989            & TN: 433 FN: 373 FP: 2 TP: 127 & 0.254                          \\
  \textbf{5}       & 0.4089      & 0.258           & 0.9847             & 0.6011            & TN: 433 FN: 371 FP: 2 TP: 129 & 0.258                          \\
  \textbf{8}       & 0.4063      & 0.256           & 0.9846             & 0.6               & TN: 433 FN: 372 FP: 2 TP: 128 & 0.256                          \\ \bottomrule
  \end{tabular}
\end{adjustbox} %this
  \end{table}

Table~\ref{Tab:table8} shows the results of testing our \senet using the same setting described in Section~\ref{sec:senetdetection1} on adversarial examples constructed with PGD at different values of $\epsilon$. 
We can see that the detection rate at 1\% false positive rate decreases sharply as $\epsilon$ grows above 0.3. At 0.5, a user may miss to notice the perturbation but the detection rate decreases to only 0.616.

\begin{comment}
At epsilon 0.1, the model yielded similar detection performance to the model without adversarial examples, with just a 0.8\% decrease in detection rate at 1\% false positive rate. However, the same score decreases by \%19.6 and \%38 percent at epsilon 0.3 and 0.5 simultaneously and worsens dramatically with higher epsilon values. 
We began adversarial training from scratch, randomly injecting adversarial examples into batches. These examples are chosen randomly from a list of three epsilon values: 0.3, 0.5, and 1. Then, we decided to adversarially train our model from zero by randomly introducing adversarial examples into batches. 
\end{comment}

\begin{figure}[!htb]
    % \centering
    % \begin{center}
        % \vspace{18pt}%
        % \caption@setup{format=plain,justification=centering}
        \includegraphics[width=\columnwidth,keepaspectratio]{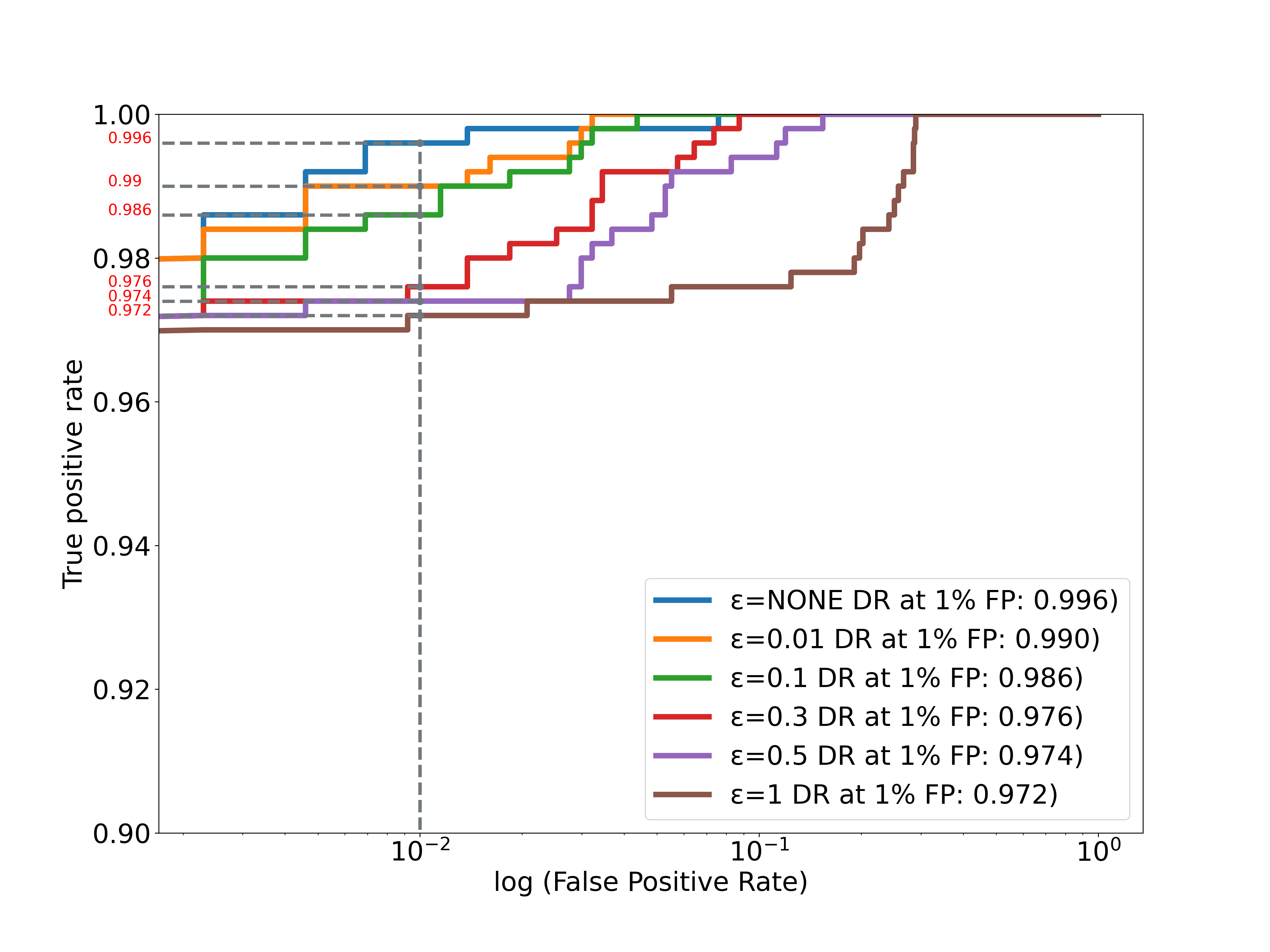}
        
        \caption{ROC Curve for adversarially trained model.}
    % \end{center}

    \label{fig:roc_PGDROC}    
  \end{figure}

\vspace{3pt}
\noindent
{\bf Adversarial Training}: To make our model more robust to adversarial examples, we follow recommendations from previous work~\cite{madry2017towards}. In practice, starting from images in the training dataset, we generate a large number of adversarial examples using PGD with different values of $\epsilon$ (specifically, 0.3, 0.5 and 1). Then, during training, we select a random amount of samples to inject into each training batch. % Although noise is very visible at $\epsilon=1$, we introduced those examples to increase the model's robustness for future attacks. 
For testing, we starts from the test dataset and generate adversarial page screenshots. We then tested \senet against these adversarial examples. The results are reported in Table~\ref{Tab:table9} and Figure~\ref{fig:roc_PGDROC}. As expected, the model becomes significantly more robust to attacks, even at $\epsilon=1$, with a detection rate above 97\% at 1\% false positives.

To further test the performance of our model on never-before-seen SE attack campaigns, we adversarially trained our model and tested it on never-before-seen attack campaigns, in a way similar to the the experiments reported in Section~\ref{sec:senetdetection3}. Notice that the model did not see examples (adversarial or not) from those attack campaigns during training.
Below, we report results for two campaigns, namely {\em Campaign 4} and {\em Campaign 10}.  The results of these new experiments can be seen in Table~\ref{Tab:table10} and Table~\ref{Tab:table11}. We can see that even in this more difficult case, adversarial examples from the two campaigns can be identified with $\approx 90$\% detection rate event at $\epsilon=1$, for which perturbations are clearly noticeable by humans, and with a detection rate of more than $95$\% for lower values of $\epsilon$, for which visual perturbations are more subtle.

\begin{table}[]
  \captionof{table}{Results after adversarial training.\label{Tab:table9}} %this
  \centering %this
\begin{adjustbox}{width=\columnwidth} %this
\small %this
  \begin{tabular}{@{}ccccccc@{}}
  \toprule
  \textbf{$\epsilon$} & \textbf{F1} & \textbf{Recall} & \textbf{Precision} & \textbf{Accuracy} & \textbf{Confusion Matrix}        & \textbf{DR at 1\% FP} \\ \midrule
  \textbf{NONE}    & 0.994       & 0.992           & 0.996              & 0.9936            & TN: 433 FN: 4 FP: 2 TP: 496  & 0.996                        \\
  \textbf{0.01}    & 0.992       & 0.99            & 0.994              & 0.9914            & TN: 432 FN: 5 FP: 3 TP: 495  & 0.99                         \\
  \textbf{0.1}     & 0.989       & 0.986           & 0.992              & 0.9882            & TN: 431 FN: 7 FP: 4 TP: 493  & 0.986                        \\
  \textbf{0.3}     & 0.9829      & 0.974           & 0.9919             & 0.9818            & TN: 431 FN: 13 FP: 4 TP: 487 & 0.976                        \\
  \textbf{0.5}     & 0.9838      & 0.974           & 0.9939             & 0.9829            & TN: 432 FN: 13 FP: 3 TP: 487 & 0.974                        \\
  \textbf{1}       & 0.9808      & 0.97            & 0.9918             & 0.9797            & TN: 431 FN: 15 FP: 4 TP: 485 & 0.972                        \\ \bottomrule
  \end{tabular}
\end{adjustbox} %this
  \end{table}

\begin{comment}
After the adversarial training, the model's detection performance increased, and the detection rate at 1 percent false positives remained above 97\% for all the epsilon values. Even when the noise is visible at epsilon 1, the detection score increased from 27\% to 97.2\%.
To further test the performance of our model on never-before-seen SE attack campaigns, we adversarialy trained our model and tested it on the fourth and tenth SE attack campaigns that can be seen in Figure~\ref{fig:MCLUSTERS}. It should be noted that the model did not see those attack campaigns during the training. The results can be seen in Table~\ref{Tab:table10} and Table~\ref{Tab:table11}
\end{comment}

\begin{table}[h]
  \captionof{table}{Results after adversarial training when SE {\em Campaign 4} is excluded from training and used for testing\label{Tab:table10}} %this
  \centering %this
\begin{adjustbox}{width=\columnwidth} %this
\small %this
  \begin{tabular}{@{}ccccccc@{}}
  \toprule
  \textbf{$\epsilon$} & \textbf{F1} & \textbf{Recall} & \textbf{Precision} & \textbf{Accuracy} & \textbf{Conf.   Matrix}    & \textbf{DR at 1\% FP} \\ \midrule
  \textbf{NONE}    & 0.9504      & 1.0             & 0.9054             & 0.9877            & TN: 493 FN: 0 FP: 7 TP: 67 & 1.0                            \\
  \textbf{0.01}    & 0.9504      & 1.0             & 0.9054             & 0.9877            & TN: 493 FN: 0 FP: 7 TP: 67 & 1.0                            \\
  \textbf{0.1}     & 0.9504      & 1.0             & 0.9054             & 0.9877            & TN: 493 FN: 0 FP: 7 TP: 67 & 1.0                            \\
  \textbf{0.3}     & 0.9504      & 1.0             & 0.9054             & 0.9877            & TN: 493 FN: 0 FP: 7 TP: 67 & 0.985                          \\
  \textbf{0.5}     & 0.9429      & 0.9851          & 0.9041             & 0.9859            & TN: 493 FN: 1 FP: 7 TP: 66 & 0.955                          \\
  \textbf{1}       & 0.9037      & 0.9104          & 0.8971             & 0.9771            & TN: 493 FN: 6 FP: 7 TP: 61 & 0.896                          \\ \bottomrule
  \end{tabular}
\end{adjustbox} %this
  \end{table}

  \begin{table}[h]
    \captionof{table}{Results after adversarial training when SE {\em Campaign 10} is excluded from training and used for testing\label{Tab:table11}} %this
  \centering %this
\begin{adjustbox}{width=\columnwidth} %this
\small %this
    \begin{tabular}{@{}ccccccc@{}}
    \toprule
    \textbf{$\epsilon$} & \textbf{F1} & \textbf{Recall} & \textbf{Precision} & \textbf{Accuracy} & \textbf{Conf.   Matrix}     & \textbf{DR at 1\% FP} \\ \midrule
    \textbf{NONE}    & 0.9832      & 1.0             & 0.967              & 0.9949            & TN: 497 FN: 0 FP: 3 TP: 88  & 1.0                            \\
    \textbf{0.01}    & 0.9832      & 1.0             & 0.967              & 0.9949            & TN: 497 FN: 0 FP: 3 TP: 88  & 1.0                            \\
    \textbf{0.1}     & 0.9832      & 1.0             & 0.967              & 0.9949            & TN: 497 FN: 0 FP: 3 TP: 88  & 1.0                            \\
    \textbf{0.3}     & 0.9659      & 0.9659          & 0.9659             & 0.9898            & TN: 497 FN: 3 FP: 3 TP: 85  & 1.0                            \\
    \textbf{0.5}     & 0.9718      & 0.9773          & 0.9663             & 0.9915            & TN: 497 FN: 2 FP: 3 TP: 86  & 0.989                          \\
    \textbf{1}       & 0.9101      & 0.8636          & 0.962              & 0.9745            & TN: 497 FN: 12 FP: 3 TP: 76 & 0.909                          \\ \bottomrule
    \end{tabular}
  \end{adjustbox} %this
    \end{table}

\begin{comment}
When epsilon 0.5 is considered the highest noise an attacker utilizes, the first model got only a 4.5\%  and the second model got a 1.1\% decrease in the detection performance. However, even in the epsilon 1 case, the same metric decreased by 10.4\% and 9.1\% respectively. 
\end{comment}

Overall, these results show that our model can be made more robust via adversarial training, even at high value of $\epsilon$ for which noise is visually noticeable and can therefore tip off the user. While we observed a slight decrease in detection rate, compared to non-adversarial test screenshots, the model still maintains a high level of accuracy and robustness to adversarial manipulations.

\subsection{\seguard Evaluation}
\label{sec:seguardeval}

As mentioned in Section~\ref{sec:seguard_setup}, VGG19 is too large of a model to be embedded into a browser extension. The main issue is represented by inference latency, which we measured at around 5 seconds per each screenshot classification. This is mostly due to the translation of the model to TensorFlow.js~\cite{tensorflowTensorFlowjsMachine} and to the large number of model parameters ($\approx$144M).

\begin{figure}[!htb]
  % \centering
  % \begin{center}
      % \vspace{18pt}%
      % \caption@setup{format=plain,justification=centering}
      \includegraphics[width=0.5\textwidth,keepaspectratio]{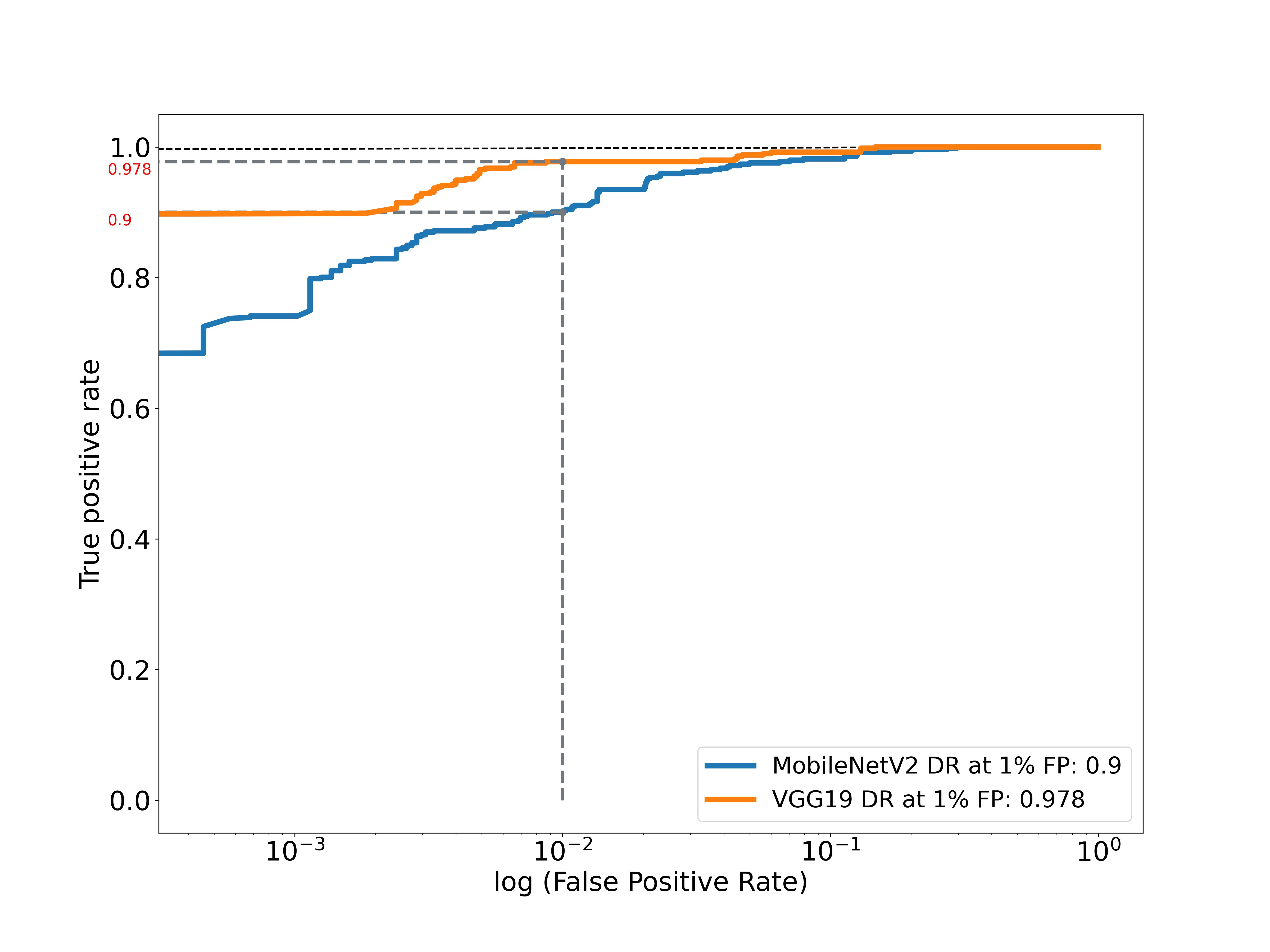}
      
      \caption{ROC curve for MobileNetV2 vs. VGG-19}
  % \end{center}

  \label{fig:roc_vggvsmob}    
\end{figure}

Unfortunately, this high inference latency is not practical for in-browser classification and for alerting the user in a timely manner. Therefore, we decided to experiment with MobileNetV2, a popular lightweight deep learning model~\cite{MobileNets}. To this end, we trained and evaluated the new model following the same setup for the VGG19 experiments we previously reported in Section~\ref{sec:sec:senetdetection2}. As shown in Figure \ref{fig:roc_vggvsmob}, MobileNetV2 perform comparably to VGG19 on our dataset, albeit with a lower detection rate at low false positive rates. For instance, at 1\% false positives both models successfully detected 90\% or more of the SE attack pages, with VGG19 outperforming MobileNetV2 by 7.8\%.
At the same time, \seguard with MobileNetV2 is able to capture, classify a page, and alert the user within $\approx$200ms. This significant reduction in latency, compared to VGG19's 5 second latency, makes in-browser SE defense more practical.

\begin{comment}
We can see that the detection rate at 1 percent false positive rate for both models is comparable; VGG19 outperformed by 7.8\%, and MobileNetV2 successfully detected 90\% of SEA campaigns at a 1\% false positive rate.
The MobileSENet, when applied, shows a detection performance similar to that of VGG19. Subsequently, we conducted a web browser latency test and discovered that the MobileSENet with the MobileNetV2 model could capture and categorize images and alert users at approximately 200ms. This significant reduction in latency, compared to VGG19's 6-second latency, greatly enhances the efficiency of our SEGuard system.
\end{comment}

\section{Discussion and Limitations}
In this section, we discuss additional considerations and limitations of our framework.

\vspace{3pt}
\noindent
\textbf{Privacy-Aware Implementation}.
As explained in Section~\ref{sec:seguard}, we implemented \seguard as a self-contained in-browser detection module by ``embedding'' the \senet model into a browser extension. To this end, we had to translate \senet into a light-weight model using MobileNet. This {\em local detection} solution is in contrast with possible alternatives based on API calls to a remote cloud-based service that can run much larger models. However, API-based detection system pose privacy risks, in that they may leak both URLs and webpage screenshots that require classification. Also, passing screenshot over the network to the API may incur significant latency. While some solutions based on locality-preserving image hashing may help to reduce privacy risks, utility and accuracy may be significantly diminished, compared to a local model.

\vspace{3pt}
\noindent
\textbf{Ethical considerations}.
Our \secrawler follows an approach for collecting SE attacks that was previously proposed in recent research~\cite{Vadrevu_IMC19,WebPushAds,TRIDENT,rafique2016s} that studied attacks delivered via malicious ads. Similar to previous works, the impact of our crawler experiments on legitimate advertisers is negligible (only few USD at most). At the same time, \secrawler allowed us to collect a large dataset of in-the-wild SE attacks and build a novel and effective in-browser defense.

\vspace{3pt}
\noindent
\textbf{Limitations}.
While on average our model is able to generalize very well to never-before-seen
screen resolutions (see Section~\ref{sec:sec:senetdetection2}), with a global
average of 97.8\% detection rate at 1\% false positive, we noticed that
the performance dropped when encountering images captured at devices of smaller
resolutions, such as 360x640 (see ROC curve in Figure~\ref{fig:ROC_mix}).
This is the smallest resolution we tested on, which is very uncommon on modern
smartphone devices\footnote{\url{https://gs.statcounter.com/screen-resolution-stats/mobile/worldwide}}. The lower performance is likely due to the fact that the
input images are scaled down and the web pages can render differently on such
small screens. With additional training examples collected for small screen sizes, it may be possible to further improve performance. We leave the study of this corner case to future work.

Also, while our system generalizes well to most never-before-seen campaigns (see Section~\ref{sec:senetdetection3}), our system may naturally struggle to detect entirely new SE attack campaigns that do not resemble anything that has ever been observed before, or in those cases in which the attack pages contain very minimal visual content (e.g., almost entirely blank pages with little text), as in the case of some notification stealing campaigns. However, it should be noticed that the vast majority of SE attack campaigns do need to include glaring visual content, to attract potential victims, and that while new campaigns and attack instances may differ from previously observed ones, they do often follow a theme and thus some resemblance with past campaigns samples, as also observed in a previous measurement study~\cite{Vadrevu_IMC19}.

Another concern is due to the fact that we train our \senet classified based on
in-the-wild SE attacks collected by \secrawler. This in-the-wild data collection
approach may open our system to potential data poisoning attacks. For instance,
attackers may try to modify their attack pages to mislead \senet when making
decision at test time. This may require a more careful manual vetting and
labeling of the data, as well as the use of measures to detect poisoning \cite{PoisonDetect_Usenix23} and mitigate its effects on the model~\cite{FLAME_Usenix,FLDetector}.
% \roberto{Karthiak, can you find a couple of appropriate references for these?}

\begin{comment}
Although our model is effective in identifying never-seen attack campaigns, attackers may introduce new attack campaigns that may evade detection by our current model. But, given that the attacks would still follow the characterisitics of the past SE attacks in terms of glaring visual cues and campaigns, our model could simply be retrained at regular intervals with new sets of training data. Since, our system already employs federated learning to train the global model, similar approach could be applied wherein the data from each user machine is used to train the model at that user machine locally and transfer its learning to the global model at frequent intervals. This further ensures the privacy of the user by only transmitting learning weights to the global model. However, owing to data corruption via poisoning attacks, it is crucial to employ additional measures to ensure the purity of the training data as proposed in studies~\cite{x}. Since this requires additional engineering effort, we leave it for future work. 
\end{comment}

\section{Related Works}
Most previous research on SE attacks focuses on specific categories of social engineering. For instance, Miramirkhani et al.~\cite{miramirkhani2017} performed an analysis of Technical Support Scams and reported websites and phone numbers used by scammers. Stone-Gross et al. ~\cite{FakeAVEconomy} showed how victims could be deceived to install fake anti-virus programs by employing scareware SE attacks. Kharraz et al. ~\cite{KharrazRK18} employed a machine learning model to detect online survey scams that led victims to reveal personal information for fake prizes or content. Our work is different because we focus on detecting an effective and practical in-browser system for detecting generic SE attack pages.

Other studies have focused primarily on identifying and measuring the occurrence of social engineering attack campaigns, without offering a detection solution. For instance, Subramani et al. ~\cite{WebPushAds} proposed a system called PushAdMiner to collect and discover web push notification messages that can deliver social malicious ad campaigns, whereas Vadrevu et al.~\cite{Vadrevu_IMC19} introduced a measurement system that automatically collects examples of SE attacks and identifies previously unknown ad networks that promote SE attack campaigns.

Recently, Yang et al.~\cite{TRIDENT} proposed a first approach towards detecting and blocking generic web-based SE attacks in the browser. The proposed system, named TRIDENT, primarily targets SE ads injected into publishers' webpages by low-reputation ad networks. SE ads are non-traditional ads that themselves utilize social engineering techniques, such as transparent overlays that perform clickjacking. TRIDENT is able {\em indirectly} detect SE attacks by identifying SE ads. However, not all SE attacks are distributed via SE ads. Unlike~\cite{TRIDENT}, our framework directly aims at detecting SE attacks and by recognizing their visual traits.

There is also a large body of research that focuses on detecting phishing attacks. While Phishing can be considered as a subclass of Social Engineering~\cite{syafitri2022social}, it is characterized by different visual traits and attack mechanisms, as we discussed in Section~\ref{sec:intro}. Recent research has focused on detecting Phishing websites using visual cues. For instance, Abdelnabi et al.~\cite{VisualPhishNet} used triplet convolutional networks to detect phishing pages by visual similarity. Lin et al. ~\cite{lin2021phishpedia} detect Phishing pages by visually detecting abused company logos. Liu et al~\cite{liu2022inferring} presented a technique that is a combination of machine learning and browser instrumentation to detect a phishing page by not only using visual cues but also discovering the intention of a phishing webpage that perform credential stealing via web forms. Unlike the above solutions, our system is able to detect generic SE attacks beyond phishing, even if no specific benign website/logo is abused or in absence of credential stealing attempts.

\begin{comment}
There is also other research that focuses on ad blocking based on visual content. Din et al. created "Percival,"~\cite{din2020percival} a powerful in-browser ad-blocker that employs deep learning to identify and block ads by analyzing perceptual features. They trained a MobilenetV2 model for this purpose.
\end{comment}

\section{Conclusion}
We presented \seshield, a framework for in-browser
detection of social engineering attacks. \seshield consists of three
main components: (i) a custom security crawler, called \secrawler, that is
dedicated to scouting the web to collect examples of in-the-wild SE attacks;
(ii) \senet, a deep learning-based image classifier trained on data collected by
\secrawler that aims to detect the often glaring visual traits of SE attack
pages; and (iii) \seguard, a proof-of-concept extension that embeds \senet into
the web browser and enables real-time SE attack detection. We performed an
extensive evaluation of our system and showed that SENet is able to  detect new
instances of SE attacks with a detection rate of up to 99.6\% at 1\% false
positive, and that is able to detect previously unseen SE attack campaigns.

\bibliographystyle{IEEEtran}
\bibliography{IEEEabrv,references_bib}

\newpage
{\bf A. ADDITIONAL DATA}

Tables~\ref{tab:ad_networks} and~\ref{Tab:imres} present additional details about our data collection. Table~\ref{tab:ad_networks} lists the ad networks we used to find seed URLs, whereas Table~\ref{Tab:imres} shows a detailed breakdown of our dataset in terms of number of images per screen resolution.

\begin{table}[h]
    \caption{Distribution of Seed URLs across Low-Tier Ad Networks}
    \label{tab:ad_networks}
    \centering
    \small
    \begin{tabular}{c|c}
        \toprule
        \hline
    Ad Network & \multicolumn{1}{c}{\begin{tabular}[c]{@{}c@{}}Seed URLs\end{tabular}}  \\ \hline 
    Adroll     & 21716                                                                                                                                                       \\ \hline
    Adblade    & 2161                                                                                                                                                            \\ \hline
    Adpushup   & 373                                                                                                                                                           \\ \hline
    Adsupply   & 205                                                                                                                                                             \\ \hline
    Pop Ads    & 178                                                                                                                                                              \\ \hline
    Adcash     & 147                                                                                                                                                            \\ \hline
    AdMaven    & 118                                                                                                                                                              \\ \hline
    Ad4Game    & 42                                                                                                                                                               \\ \hline
    AdReactor  & 24                                                                                                                                                                 \\ \hline
    Adsense    & 15                                                                                   
    \\ \hline 
    \bottomrule        
    \end{tabular}
\end{table}

\begin{table}[h]
    \captionof{table}{The Number of SE attacks and benign page samples per screen resolution}
    \centering %this
    % \begin{adjustbox}{width=\columnwidth} %this
    \small %this
    \begin{tabular}{@{}ccc@{}}
    \toprule
    \textbf{Resolution} & \textbf{\# Benign pages} & \textbf{\# SE attacks} \\ \midrule
    \textbf{360x640}    & 2402                             & 255                                 \\
    \textbf{360x740}    & -                                & 1481                                \\
    \textbf{414x896}    & 3398                             & 24                                  \\
    \textbf{750x1334}   & 7979                             & -                                   \\
    \textbf{768x1024}   & 5067                             & 51                                  \\
    \textbf{800x1280}   & 4657                             & 61                                  \\
    \textbf{1024x768}   & 4595                             & 21                                  \\
    \textbf{1280x800}   & 4162                             & 36                                  \\
    \textbf{1200x803}   & 8067                             & -                                   \\
    \textbf{1366x677}   & 263                              & 11                                  \\
    \textbf{1366x720}   & 1102                             & 18                                  \\
    \textbf{1366x724}   & 268                              & 6                                   \\
    \textbf{1366x728}   & 14530                            & 357                                 \\
    \textbf{1366x738}   & 1137                             & 22                                  \\
    \textbf{1366x741}   & 498                              & 9                                   \\
    \textbf{1366x768}   & 1809                             & 54                                  \\
    \textbf{1478x837}   & 757                              & 34                                  \\
    \textbf{1536x816}   & 1364                             & 65                                  \\
    \textbf{1536x824}   & 14432                            & 474                                 \\
    \textbf{1536x826}   & 776                              & 25                                  \\
    \textbf{1536x834}   & 2063                             & 84                                  \\
    \textbf{1536x864}   & 778                              & 25                                  \\
    \textbf{1785x993}   & -                                & 3267                                \\
    \textbf{1858x1053}  & 869                              & 30                                  \\
    \textbf{1858x1080}  & 813                              & 21                                  \\
    \textbf{1920x998}   & 1960                             & 41                                  \\
    \textbf{1920x1032}  & 1721                             & 51                                  \\
    \textbf{1920x1040}  & 16909                            & 502                                 \\
    \textbf{1920x1050}  & 2354                             & 73                                  \\
    \textbf{1920x1052}  & 829                              & 19                                  \\
    \textbf{1920x1080}  & 3195                             & 325                                 \\
    \textbf{1920x1097}  & 1126                             & 42                                  \\ \bottomrule
    \end{tabular}
    % \end{adjustbox} %this
   \label{Tab:imres}
    \end{table}

\end{document}